\documentclass[UTF8,a4paper]{article}
\usepackage{amsmath}
\usepackage{graphicx}
\usepackage{subfigure}
\usepackage{epstopdf}
\usepackage{geometry}
\usepackage{ulem}
\usepackage{indentfirst}
\usepackage{amsfonts,amssymb,dsfont}
\usepackage{setspace}
\usepackage{threeparttable}
\usepackage{amsmath,mathtools,amsthm}
\usepackage{array}
\usepackage{extarrows}
\usepackage{upgreek}

\makeatletter
\renewcommand*\env@matrix[1][\arraystretch]{%
  \edef\arraystretch{#1}%
  \hskip -\arraycolsep
  \let\@ifnextchar\new@ifnextchar
  \array{*\c@MaxMatrixCols c}}
\makeatother
\begin{document}


\title{\bf Electronic properties of the parabolic Dirac system}
\author{Chen-Huan Wu
\thanks{chenhuanwu1@gmail.com}
\\College of Physics and Electronic Engineering, Northwest Normal University, Lanzhou 730070, China}

\maketitle
\vspace{-30pt}
\begin{abstract}
\begin{large} 

We investigate the electronic properties of the parabolic Dirac system.
In the presence of magnetic field,
we discuss the thermodynamical potential and the anomalous diamagnetism due to the presence of Dirac cone.
The integer Hall conductivity and the intrinsic Hall conductivity induced by the Berry curvature are also investigated.
Considering the disorder-induced self-energy in self-consistent Born approximation,
the RKKY interaction, long-range scattering rate and the spectral function
are also calculated,
which are distinct to that of the clean system (without impurity).
\\

\end{large}

\end{abstract}
\begin{large}

\section{Introduction}

The 2D topological Dirac systems,
which with tunable the bandgap and realizable band inversion
(like the anomalous Hall effect),
have a huge potential in the spintronics and pseudospintronics.
For the doped (extrinsic) or multilayer 2D Dirac system, the linear dispersion 
near the Dirac point is replaced by the parabolic dispersion,
which is still topological non-trivial.
Distinct to the parabolic metal,
the Landau levels (LLs) of the parabolic Dirac system is non-equispaced
due to the existence of the Dirac node,
and disperse parabolically along the direction of the magnetic field.
The LLs in parabolic system is also related to the strength of the trigonal warping\cite{McCann E}.
Thus all of the filled LLs contribute to the Anomalous diamagnetism
as described by the Wolff Hamiltonian under a weak magnetic field.
The 3D Dirac or Weyl semimetal have the same effect due to the exisence of the Dirac node or Weyl node within the bulk.
Similar to the Heusler compounds where the topological phase transition can be realized by applying the compression
which can change the strength of the spin-orbit couping (SOC) as well as the lattice constant\cite{Chadov S},
the strain is also valid in 2D Dirac system in realizing the topological phase transition\cite{Wan W,Teshome T,Teshome T2}.
The quantum Hall effect and the $Z_{2}$ topological invariants characterized by Wilson loop\cite{Yu R},
are play the important role in teh search for the novel 2D materials by the high throughput screening\cite{Li X}.

The electronic transport of the Dirac or Weyl systems have arouse widely theoretic and experimental explorations\cite{Haratipour N}.
We investigate the properties of the parabolic Dirac system as well as the multilayer bulk form.
Firstly, unlike the linear Dirac system where the momentum is coupled with the degrees of freedom described by 
the Pauli matrix,
the momentum of the parabolic system does not related to the Pauli matrix except for the anisotropic dispersion.
And that's also different to the three-dimensional Dirac or Weyl system\cite{Ahn S}.
Under magnetic field,
the thermodynamical potential, Hall conductivity, and the orbital diamagnetic susceptibility in the presence of impurity-induced
disorder are studied.
Base on the self-energy-corrected Green's function in real space,
the RKKY interaction is also discussed, as well as the important range function of the multilayer form.
We mainly focus on the disorder-induced self-energy,
which contribute by both the random phase approximation (RPA)-type contribution and the non-RPA-type contribution
as clear seem in the diagramatic representations of this paper.
The spectral function of the parabolc system is also discussed.

\section{Model}

For the bilayer 2D Dirac system,
we take the bilayer silicene as an example whose
low-energy effective Hamiltonian reads
\begin{equation} 
\begin{aligned}
H^{bi}=
\tau'\cdot H-\frac{t'}{2}(\tau'_{x}\tau_{x}-\tau'_{y}\tau_{y})+\frac{t_{w}}{2}k_{x}(\tau'_{x}\tau_{x}+\tau'_{y}\tau_{y})
+\lambda'_{SOC}\tau_{z}(\tau'_{y}s_{x}+\eta\tau'_{x}s_{y}),
\end{aligned}
\end{equation}
where $t_{w}$ is the interlayer hopping which related to the trigonal warping, and the value of 
$t_{w}$ decreases with the increase of layer number,
e.g., it's much smaller than the intralayer nearest hopping $t$ in the bulk Dirac materials.
with the chiral quasiparticles with 2$\pi$ Berry phase in pseudospin space when massless,
and here the monolayer Hamiltonian is
\begin{equation} 
\begin{aligned}
H=\eta(Ds_{z}\tau_{z}-\frac{({\bf k}\cdot{\pmb \tau})^{2}}{2m})-\mu,
\end{aligned}
\end{equation}
where $\eta=\pm 1$ denotes the valley degree of freedom, $s_{z}$ and ${\pmb \tau}$ denote the spin and pseudospin degrees of freedom, respectively.
$D$ is the Dirac mass, $\mu=v_{F}k_{F}$ is the chemical potential.
The eigenenergies can be obtained through solving the above Hamiltonian, which reads
\begin{equation} 
\begin{aligned}
\varepsilon_{\pm}=\frac{-2m\mu\pm \sqrt{k^{4}+4m^{2}(D^{2}+\mu^{2})+4\eta k^{2}m\mu{\rm cos}2\Phi}}{2m},
\end{aligned}
\end{equation}
where $m=0.268m_{0}$ is the effective mass, $\phi={\rm atan}k_{y}/k_{x}$. For isotropic dispersion,
$\Phi=\pi/4$,
and thus the eigenenergies reduced to 
\begin{equation} 
\begin{aligned}
\varepsilon_{\pm}=\frac{-2m\mu\pm \sqrt{k^{4}+4m^{2}(D^{2}+\mu^{2})}}{2m}.
\end{aligned}
\end{equation}

For the case of anisotropic dispersion,
\begin{equation} 
\begin{aligned}
H=
\begin{pmatrix}[1.5]
D\tau_{z}+a_{x}k_{x}^{2}+a_{y}k_{y}^{2}&-\frac{(\eta k_{x}-ik_{y})^{2}}{2m^{*}}\\
-\frac{(\eta k_{x}+ik_{y})^{2}}{2m^{*}}&-D\tau_{z}-b_{x}k_{x}^{2}-b_{y}k_{y}^{2},
\end{pmatrix}-\mu
\end{aligned}
\end{equation}
where we drop the real spin degree of freedom here since it can be incorporated into the Dirac-mass term.
Then the eigenenergies can be obtained as
\begin{equation} 
\begin{aligned}
\varepsilon_{\pm}=&\frac{1}{2m}(a_{x}k_{x}^{2}m-b_{x}k_{x}^{2}m+a_{y}k_{y}^{2}m-b_{y}k_{y}^{2}m-2m\mu\mp\sqrt{F}),\\
F=&k_{x}^{4}(1+a_{x}^{2}m^{2}+2a_{x}b_{x}m^{2}+b_{x}^{2}m^{2})+k_{y}^{4}(1+a_{y}^{2}m^{2}+2a_{y}b_{y}m^{2}+b_{y}^{2}m^{2})\\
&+4k_{y}^{2}m(a_{y}Dm+b_{y}Dm-\mu)+4m^{2}(D^{2}+\mu^{2})+2k_{x}^{2}(k_{y}^{2}(1+a_{y}b_{x}m^{2}+a_{x}(a_{y}+b_{y})m^{2})\\
&+2m(a_{x}Dm+b_{x}Dm+\mu)).
\end{aligned}
\end{equation}
The disorders, including the local charged impurity or vacancy,
will lead to the instability as well as the large density of states (DOS) at Dirac-point.
The increasement of DOS at Fermi level also results from the LLs as introduced by the artificial gauge field.
For a 2D Dirac system exposed to the artificial gauge field,
the groud Fock state can be expressed as $|\psi_{0}\rangle=\Pi_{\nu,j}c_{\nu,j}^{\dag}|0\rangle$ in the presence of SU(4) symmetry
with a small lattice spacing $a=\Lambda^{-1}$,
where $\nu\ge 1$ is the filling factor (integer) gives rise by the Landau quantization and $|0\rangle$ is the vacuum state,
$j$ is the total angular momentum consist of the spin and orbital angular momentum mixed by the strong SOC.
The four fold degenerate LLs is possible when the Dirac-mass $D=0$.
By the minimal substitution $k\rightarrow k-e{\bf A}$ where $e$ is the synthetic charge and 
${\bf A}$ is the vector potential determined by the Zeeman shift,
we have 
$[k_{x},k_{y}]=1/i\ell_{B}^{2}$ and
$[k_{+},k_{-}]=\frac{2\hbar^{2}}{\ell_{B}^{2}}=2\hbar eB$
where $\ell_{B}$ is the magnetic length.
The vector potential here is direction-dependent except for the case of Lorentz invariant.
That's in distinct from the superconducting case, where the quasiparticles doesn't minimally couple to the ${\bf A}$
but to the supercurrent\cite{Ghosal A}.
Due to the gauge invariance, the vector potential $A$ has the same scaling dimension with the momentum
which is $1/S$ here and $S$ is the area of unit cell.
For gapless case, the degeneracy number is $g\ell_{B}^{-2}/2\pi S$ in $n=0$ LL and $2g\ell_{B}^{-2}/2\pi S$ in $n \neq 0$ LLs. 
In contrast to the metallic phase,
the semimetallic phase is more stable against the disorder due to the less electronic instability at the charge neutrality.

By coupling the orbital degree of freedom to the magnetic field,
the orbital term (in contrast to the kinetic term and the Zeeman term) gives rise to the Landau diamagnetism.
Unlike the free electron,
the Dirac electron acts a $\sqrt{B}$ (or $\sqrt{k_{z}}$) behavior respect to the total angular momentum,
that can also be proved by the eigenvalue of the Hamiltonian under a magnetic field
which reads, in a simplest form, $\varepsilon_{n}^{\pm}=\pm\sqrt{2n\hbar^{2}v_{F}^{2}/\ell_{B}^{2}+D^{2}}$ for $n\neq 0$ levels
and $\varepsilon_{0}=D$ for $n=0$ level,
where the sign $\pm$ correspond to electron/hole
and can be exchanged by the vertex function in electron-hole channel.
The above eigenenergies of the LLs can be obtained by solving Hamiltonian (under a perpendicular magnetic field)\cite{Shakouri K,McCann E}
\begin{equation} 
\begin{aligned}
H=
\begin{pmatrix}[1.5]
D&\hbar v_{F}(k_{x}+\eta\partial_{y}-y/\ell_{B}^{2})\\
\hbar v_{F}(k_{x}-\eta\partial_{y}-y/\ell_{B}^{2})&-D,
\end{pmatrix}-\mu.
\end{aligned}
\end{equation}
In the case of time-reversal invariant,
the Kramers degeneracy is lifted by the magnetic field, and then the eigenenergies of LLs becomes
$\varepsilon_{n}^{\pm}=\pm\sqrt{(2n-1+\sigma_{z}/2)\hbar^{2}v_{F}^{2}/\ell_{B}^{2}+D^{2}}$
where $\sigma_{z}=\pm 1$ is the spin projection on the magnetic field
which related to the Zeeman splitting when the the total spin is nonzero\cite{Park J}.
The $\sigma_{z}$ here is a good quantum number in the presence of time revesal invariant (TRI)
and the absence of Rashba coupling.
The $\sigma_{z}$-dependent Zeeman splitting is usually much weaker than the LL splitting in the 2D Dirac system,
which reads $\sigma_{z}g\mu_{B}B$ with $\mu_{B}=\hbar e/2mc$ the Bohr magneton,
however, for the 3D Dirac electrons in the non-relativistic limit\cite{Yang K},
like in the context of
Bismuth\cite{Fuseya Y}, the Landau level splitting could be equals to the Zeeman splitting
due to the large longitudinal effective mass which results in a large $g$-factor.
Here the $g$-factor reads $g=2m/m_{c}\gtrsim  2$ where 
$m^{-1}_{c}=\sqrt{\frac{2c}{\hbar eB}}v_{F}\propto \frac{v_{F}^{2}}{D}$ (in ultrarelativistic limit)
is the cyclotron effective mass.
Both the magnetic field-induced symmetry breaking (as well as the magnetization) and the $m_{c}$ is proportional to the $\sqrt{B}$.
When the Dirac mass is small enough (and thus the interband transition is rised) or when the Dirca particle is ultrarelativistic,
the $g$-factor would be very large.
Besides, the effective mass is also affected by the temperature in non-adiabatical case,
with the increase of temperature,
the effective mass will decrease and thus the $g$-factor is enhanced, e.g.,
in
the GaAs\cite{Hubner J},
the effective mass the $g$-factor have the variation pattern as shown in the Fig.1.
For the QED free electron in ultrarelativistic case,
the eigenenergies becomes
$\varepsilon_{n}^{\pm}=\pm\sqrt{(2n+1+\sigma_{z})\hbar^{2}v_{F}^{2}/\ell_{B}^{2}+D^{2}}$\cite{Orlita M,Fuseya Y,Koshino M}.

The linear LLs is a signature of the free electron in $B\rightarrow 0$ limit\cite{Fuseya Y},
while for the bulk Dirac system, the linear LLs and the parabolic LL are coexist,
as found in the topological insulator $ {\rm Bi}_{ 0.91}{\rm Sb}_{ 0.09 }$\cite{Schafgans A A}
where the intra LL transition is dominating.
While for another kind of bulk materials with the bulk Dirac cone as found in 3D Dirac semimetal,
like the Cd$_{3}$As$_{2}$\cite{Neupane M} and Na$_{3}$Bi\cite{Liu Z K},
the LL spacing are non-uniform too,
and with the parabolic dispersion.
For the single Dirac cone exists in the surface state of the Bi$_{2}$Se$_{3}$ or the 3D narrow gap semiconductor like the
Hg$_{1-x}$Cd$_{x}$Te,
the LL are also parabolic except at very low magnetic field ($B<0.01$ T)\cite{Orlita M},
and although the Dirac Fermions here is distinct from the massless particles in QED (ultrarelativistic)
since their excited velocity is much slower that the speed of light,
they have some common points in the aspect of electronic transport,
some common features are stated in the Ref.\cite{Kashuba A B} for the ideal defectless graphene.
Under strong magnetic field, the giant diamagnetism contributed by all of the filled states 
(for Dirac bands) diverges logarithmically when the chemical potential is close to the Dirac cone,
for a discussion about the anomalous divergence of the diamagnetism susceptibility
near the Dirac node at low-temperature, see, e.g., Refs.\cite{Wu C Hcurrent}.

\section{Thermodynamical potential in clean system under magnetic field}

The noninteracting thermodynamical potential read
\begin{equation} 
\begin{aligned}
\Omega=-T\int^{\infty}_{-\infty}dE D(E){\rm ln}(2{\rm cosh}\frac{E-\mu}{2T}),
\end{aligned}
\end{equation}
where $T$ is the absolute temperature,
here the integral is over both the occupied states and unoccupied states since the
thermal energy will excites the electrons into arbitary states.
While for zero-temperature case,
the maximal LL is $N={\rm Int}[\frac{\mu^{2}c}{2\hbar ev_{F}^{2}B}]$
where ${\rm Int}[z]$ denotes the integer part.
That's to say, even at zero-temperature,
both the free energy and thermodynamical potential are nonzero.
The DOS $D(E)$ which is dramatically enhanced by the LL effect reads
\begin{equation} 
\begin{aligned}
D(E)=&\frac{g}{2\pi\ell_{B}^{2}}\left[\int\frac{d^{2}k}{(2\pi)^{2}}\delta(E-\varepsilon(k))\right.\\
&\left.+\sum^{N}_{n=1}(\delta(E-\varepsilon_{n})+\delta(E+\varepsilon_{n}))\right],
\end{aligned}
\end{equation}
where $N$ is the maximal LL index and the sum is over the $n\neq 0$ levels since only the $n\neq 0$ levels contribute to the diamagnetism.
For strong enough magnetic field where the peak of $B$-dependent DOS are separed and the 
number of states in this case is $g/2\pi\ell_{B}^{2}$ per LL\cite{Shakouri K}. 
Here the eigenenergies $\varepsilon(k)$ is contained since we apply a weak magnetic field.
Near Dirac point where we assume simply the linear relation, the first term of the thermodynamical can be written as
\begin{equation} 
\begin{aligned}
\Omega_{1}=&-T\int^{\infty}_{-\infty}dE\frac{g}{2\pi\ell_{B}^{2}}\int\frac{d^{2}k}{(2\pi)^{2}}\delta(E-\varepsilon(k)){\rm ln}(2{\rm cosh}\frac{E-\mu}{2T})\\
=&-T\int^{\infty}_{-\infty}dE\frac{1}{S}\frac{g}{2\pi\ell_{B}^{2}}\int^{\pi}_{0}d\theta\int^{\Lambda}_{0}kdk\delta(E-k){\rm ln}(2{\rm cosh}\frac{E-\mu}{2T})\\
=&-T\int^{\infty}_{-\infty}dE\frac{\pi}{S}\frac{g}{2\pi\ell_{B}^{2}}E\theta(E)\theta(\Lambda-E){\rm ln}(2{\rm cosh}\frac{E-\mu}{2T})\\
=&-T\frac{\pi}{S}\frac{g}{2\pi\ell_{B}^{2}}
 \theta(E)[T^{2}{\rm Li}_{3}(-e^{(\mu-E)/T})\\
&+ET{\rm Li}_{2}(-e^{(\mu-E)/T})+T^{2}(-{\rm Li}_{3}(-e^{\mu/T})-\frac{1}{12T}(E-\mu)\\
&\times (-2\mu^{2}+E^{2}+\mu E+6T(\mu+E){\rm ln}(e^{(\mu-E)/T}+1)-6T(\mu+E){\rm ln}(2{\rm cosh}\frac{E-\mu}{2T}))\\
&+\frac{1}{6T}\mu^{2}(\mu-3T{\rm ln}(e^{\mu/T}+1)+3T{\rm ln}(2{\rm cosh}\frac{\mu}{2T}))]\bigg|_{E},
\end{aligned}
\end{equation}
where ${\rm Li}_{n}(z)=\sum^{\infty}_{k=1}\frac{z^{k}}{k^{n}}$ is the polylogarithm and here
we assume $\Lambda>E>0$ and thus the range of $E$ can't take the whole real space,
and the second term reads
\begin{equation} 
\begin{aligned}
\Omega_{2}=&-T\int^{\infty}_{-\infty}dE \frac{g}{2\pi\ell_{B}^{2}}\sum^{N}_{n=1}(\delta(E-\varepsilon_{n})+\delta(E+\varepsilon_{n}))
 {\rm ln}(2{\rm cosh}\frac{E-\mu}{2T})\\
\approx &-T\frac{g}{2\pi\ell_{B}^{2}}
[{\rm ln}({\rm cosh}\frac{\mu-\sqrt{N}}{2T})+{\rm ln}({\rm cosh}\frac{\mu-\sqrt{(N-1)}}{2T})+\cdot\cdot\cdot+\\
&{\rm ln}({\rm cosh}\frac{\mu-1}{2T})+{\rm ln}({\rm cosh}\frac{\mu+1}{2T})+\cdot\cdot\cdot+
{\rm ln}({\rm cosh}\frac{\mu+\sqrt{N}}{2T})+{\rm ln}(4^{N})],
\end{aligned}
\end{equation}
here we assume the maximal energy $E(<\Lambda)$ is larger than the maximal LL index $N$.
Then the thermodynamical potential is $\Omega=\Omega_{1}+\Omega_{2}$,
which is presented in the Fig.2 as a function of temperature.
In Fig.2, we set the chemical potential as 0.02 eV,
and we can see that the LLs term $\Omega_{2}$ reduce the thermodynamical potential in contrast to the $\Omega_{1}$
we also show the $\Omega_{1}$ at zero-energy case in the inset,
which we can see a dramatic oscillation 

Note that the above results are only valid for the case of weak magnetic field,
while for the strong magnetic field,
the thermodynamical potential contributed only by the eigenenergies $\varepsilon(k)$ where $k=n$,
i.e., the momentum of the corresponding LL.
Further, a out-of-plane high magnetic field (e.g., $>45$ T\cite{Moll P J W})
may splits the Dirac node into two opposite chirality Weyl nodes
and gap out the edge state with a gap $0.2$ meV/T\cite{Zhou B,WuElectronic transport},
while a in-plane strong magnetic field won't splits the Dirac node but will gap out the edge state or the $n=0$ LL.

For linear Dirca system in the gapless case and half-filling,
the DOS reduced to 
\begin{equation} 
\begin{aligned}
D(E)=&\frac{g}{2\pi\ell_{B}^{2}}\left[\int\frac{d^{2}k}{(2\pi)^{2}}\delta(E-\hbar v_{F}k)\right.\\
&\left.+\sum^{N}_{n=1}(\delta(E-\sqrt{2n}\hbar v_{F}/\ell_{B})+\delta(E+\sqrt{2n}\hbar v_{F}/\ell_{B}))\right]\\
=&\frac{g}{2\pi\ell_{B}^{2}}\left[\pi \frac{E}{\hbar^{2} v_{F}^{2}} \right.\\
&\left.+\sum^{N}_{n=1}(\delta(E-\sqrt{2n}\hbar v_{F}/\ell_{B})+\delta(E+\sqrt{2n}\hbar v_{F}/\ell_{B}))\right],
\end{aligned}
\end{equation}
here we use the model of the gapless (or weakly gapped) parabolic model.
By setting the ultraviolet cutoff as $\Lambda=4.8$ eV,
we present the local DOS and LLs of the linear and parabolic Dirac system in Fig.3.
The sharp $\delta$-function in the second term of DOS will be broadened by the disorder come from the impurity or crack, vacancy,
by we imaging the clean simple here.
In Fig.3(a),
we don't consider the broadening of each peak, and
only the $n\neq 0$ levels contribute to the LDOS.
The variation of their spacing is related to the LLs dispersion,
at a fixed value of $B$, if the spacing between each LL is equal,
then the spacing between each DOS peak is also equal, and vice versa.
In the presence of disorder, the peaks of LDOS shown in Fig.3(a) will be broadened by the quasiparticle scattering
which is similar to the broaden of LLs.
Considering the broaden effect, the DOS can be expressed similar to the spectral function:
\begin{equation} 
\begin{aligned}
D(E)=\frac{1}{S\pi}{\rm Im}\sum_{n}^{N}\sum_{s_{z},\eta=\pm 1}\prod_{i=n,s_{z},\eta}\frac{1}{E+i\eta-E_{i}},
\end{aligned}
\end{equation}
here the term $\prod_{i=n,s_{z},\eta}\frac{1}{E+i\eta-E_{i}}$ can be decomposed into the partial fractions by the similar way used in Ref.\cite{Saito R},
and the resulting DOS is shown in Refs.\cite{Shakouri K,Ando T}.
The broadening of the LLs due to the impurity scattering reads 
$\frac{\sqrt{2}\hbar v_{F}}{\ell_{B}}\sqrt{\frac{2U_{k,k'}}{\pi v_{F}^{2}}(1+\delta_{n0})}$
which is much smaller than the LLs spacing at low magnetic field.
Here $U_{k,k'}$ is the electron-impurity interaction.
Under weak magnetic field,
the Fermi energy as a function of $B$ is related to the DOS
and behaviors a step-like action with the increase of magnetic field
and finally reduced to zero at the zeroth LL\cite{Shakouri K,Ando T,Scriba J,Shakouri K2}.
While for the parabolic system,
the DOS becomes
\begin{equation} 
\begin{aligned}
D(E)=&\frac{g}{2\pi\ell_{B}^{2}}\left[\int\frac{d^{2}k}{(2\pi)^{2}}\delta(E-\frac{k^{2}}{2m})\right.\\
&\left.+\sum^{N}_{n=1}(\delta(E-\sqrt{2n(n+1)}\hbar v_{F}/\ell_{B})+\delta(E+\sqrt{2n(n+1)}\hbar v_{F}/\ell_{B}))\right]\\
=&\frac{g}{2\pi\ell_{B}^{2}}\left[\pi m\theta(\frac{\Lambda^{2}}{m}-2E)\right.\\
&\left.+\sum^{N}_{n=1}(\delta(E-\sqrt{2n(n+1)}\hbar v_{F}/\ell_{B})+\delta(E+\sqrt{2n(n+1)}\hbar v_{F}/\ell_{B}))\right],
\end{aligned}
\end{equation}
through above expression,
we also see that the first term of DOS is independent of the band dispersion in the limit of $\Lambda\gg E$.
Now we can see that at a fixed value of $B$, the spacing between each LL is equal,
then the spacing between each DOS peak is also equal, which is agree with the above result.
However, note that the LLs of Dirac electron is always different to the free electron like in 
the non-parabolic quantum well\cite{Scriba J,Sigg H}.

\section{Conductivity and orbital susceptibility}

Considering the impurity effect,
the Hall conductivity and the intra-LLs longitudinal conductivity can be written as\cite{Wu C Hinteger}
\begin{equation}
\begin{aligned}
\sigma_{xy}=&\frac{i\hbar e^{2}}{\omega_{c}S_{B}}\sum_{E_{n'}>E_{n},s_{z},\eta}[f(E_{n})-f(E_{n'})]
\frac{\langle n|v_{x}|n'\rangle\langle n'|v_{y}|n\rangle}
{(E_{n}-E_{n'})(E_{n}-E_{n'}+\omega+i\delta+i\Gamma)},\\
\sigma_{xx}=&\frac{4\pi^{2} e^{2}\Gamma}{TS_{B}h\omega_{c}}\sum_{n,s_{z},\eta} F_{ss'}
E_{n}f(E_{n})(1-f(E_{n})\delta_{E_{n'},E_{n}},\\
\end{aligned}
\end{equation}
where we define $S_{B}=\ell_{B}^{2}k^{2}$ and the normalized velocity matrix elememts are
\begin{equation}
\begin{aligned}
\langle n|v_{x}|n'\rangle=\frac{v_{F}}{2}       (s(1+\frac{D^{++}}{E_{n}})^{1/2}(1-\frac{D^{+-}}{E_{n'}})^{1/2})\delta_{s_{z},s_{z}'}\delta_{t_{z},t_{z}'}(\delta_{n',n-1}+\delta_{n',n+1}),\\
\langle n'|v_{y}|n\rangle=\frac{i\eta v_{F}}{2}(-s(1-\frac{D^{+-}}{E_{n'}})^{1/2}(1-\frac{D^{++}}{E_{n}})^{1/2})\delta_{s_{z},s_{z}'}\delta_{t_{z},t_{z}'}(\delta_{n',n-1}-\delta_{n',n+1}),\\
\end{aligned}
\end{equation}
where $s$ the band index,
the first and second superscript of Dirac-mass denote the valley and spin indices.
The matrix element $F_{ss'}$ reads
\begin{equation} 
\begin{aligned}
F_{ss'}
=\frac{1}{4}\left[(2n+1)(1+\frac{D }{E_{n}})^{4}-2n(1+\frac{D }{E_{n}})^{2}(1-\frac{D }{E_{n}})^{2}
+(2n-1)(1-\frac{D }{E_{n}})^{4}\right],
\end{aligned}
\end{equation}
and $\Gamma=\pi n_{i}V^{2}_{0}/\hbar S_{B}$.
We plot the Hall conductivity and the longitudinal conductivity as a function of the Fermi energy in Fig.4 with the fixed
magnetic field B=1T and temperature T=0.1 K.
In the case of zero electric field\cite{Tahir M},
the Hall conductivity $\sigma_{xy}$ is quantitative as $\sigma_{xy}=2(2n+1)\frac{e^{2}}{h}$.
In the presence of broekn inversion symmetry and the electromagnetic field,
the anomalous Hall conductivity is rised by the Berry curvature
as
\begin{equation} 
\begin{aligned}
\sigma_{xy}=\frac{e^{2}}{\hbar}\int\frac{d^{2}k}{(2\pi)^{2}}N_{F}(\varepsilon_{k})\Omega_{z}(k),
\end{aligned}
\end{equation}
where the Berry curvature reads\cite{WuElectronic transport}
\begin{equation} 
\begin{aligned}
\Omega_{z}(k)
=&-{\rm Im}\left[ \sum_{\psi'\neq\psi}\frac{\langle\psi'|\partial_{k\mu}H_{k}|\psi\rangle\times\langle\psi'|\partial_{k\nu}H_{k}|\psi\rangle}{(\varepsilon_{\psi}-\varepsilon_{\psi'})^{2}}\right]
=&\eta\hbar^{2}v_{F}^{2}D/(2\varepsilon^{3/2}).
\end{aligned}
\end{equation}
After performing the integral over momentum and using the energy cutoff $\Lambda=4.8$ eV,
we obtain the intrinsic contribution from the Berry curvature as
\begin{equation} 
\begin{aligned}
\sigma_{xy}= \frac{e^{2}}{\hbar}\frac{1}{4\pi}\frac{\eta\hbar^{2}v_{F}^{2}D}{2}
\frac{0.19e^{\mu}k^{2}}{(e^{\mu}+e^{1.86\sqrt{0.04+1.072\mu^{2}}})(0.04+1.072\mu^{2})^{0.75}}+O(k^{6}).
\end{aligned}
\end{equation}
The result is presented in the Fig.5,
where we set the Dirac-mass $D=0.1$ eV,
and for case that the
Fermi level lies within the band gap, the Hall conductivity dependents only on natural constant and
the sign of Dirac-mass\cite{Tahir M2,Vargiamidis V}.

Different to the zero temperature case\cite{Thakur A},
the diamagnetic susceptibility at finite temperature reads
\begin{equation} 
\begin{aligned}
\chi=-\frac{\partial ^{2}\Omega}{\partial B^{2}}\bigg|_{B\rightarrow 0}.
\end{aligned}
\end{equation}
Since in the weak magnetic field limit,
the DOS preserve only the first term, and the summation over the LLs can be replaced by the interal over the range $-\Lambda<\varepsilon<\Lambda$.
By using the Euler-Maclaurin formula, the DOS becomes
\begin{equation} 
\begin{aligned}
D(E)=&\frac{g}{4\pi^{2}\hbar^{2}v_{F}^{2}}\left[\int\frac{d^{2}k}{(2\pi)^{2}}\delta(E-\varepsilon(k))+O(B^{6})
\right].
\end{aligned}
\end{equation}
The orbital susceptibility is very large at low-energy limit\cite{Wu C Hcurrent} 
(especially for the linear Dirac system which contains a $\delta$-term $\delta(\varepsilon)$
while for the parabolic Dirac system it's replaced by a logarithmic term ${\rm ln}\frac{\Lambda}{|\varepsilon|}$).
That reveals that the zero-energy LL contributes to the singular of susceptibility,
and in massless case it's equally shared by the conduction band and valence band.
The Pauli paramagnetism is given by the Zeeman term in spin (or pseudospin) space,
and it's smaller than the Landau diamagnetism given by the Orbital term.
Both the diamagnetic and paramagnetic response which with opposite magnetic moment (i.e.,
diamagnetic moment and paramagnetic moment with the spin carriers along the edge direction
carriers the up- and down- spin, respectively) can be coexist in the 
2D Dirac system (like the silicene\cite{Wu C HTight,Xu N})
due to the interactions
between the magnetic field and the charge carriers with spin-up and spin-down, respectively,
and they are both increse with the temperature.
In the presence of short-range disorder,
the orbital susceptibility has been obtained as\cite{Koshino M2}
\begin{equation} 
\begin{aligned}
\chi=-\frac{g_{s}g_{v}e^{2}v^{2}}{6\pi c^{2}}\int^{\infty}_{-\infty}d\Omega N_{F}(\Omega){\rm Im}\frac{1}{\Omega-\Sigma^{D}(\Omega)},
\end{aligned}
\end{equation}
where $N_{F}(\Omega)$ is the Dirac-Fermi distribution function 
and $\Sigma^{D}(\Omega)$ is the momentum- and magnetic field-independent self-energy induced by the disorder as calculated in the following section.
In zero-temperature limit,
the Dirac-Fermi distribution function can be approximately viewed as 1 for the electron occupied part,
then the orbital diamagnetic susceptibility can be obtained as 3.1236($-\frac{g_{s}g_{v}e^{2}v^{2}}{6\pi c^{2}}$) by solving the above integral
where we set $m=0.268m_{0}$, $\eta=0.1\hbar\omega_{c}$,
and $\Lambda$ equals to twice of the $\pi$-band width (i.e., nearly 4.8 eV).
In Fig.6, we plot the above orbital susceptibility under different energy cutoff.
We can find that in high energy cutoff region,
the susceptibility is logarithmic with $\Lambda$.
For the case of long-range disorder,
the susceptibility contains a Dirac $\delta$-term in the limit of impurity scattering potential,
as we explored\cite{Wu C Hcurrent}.

\section{Disorder effect and the RKKY interaction}

The self-consistently embedded problem emerges in the presence of disorder.
In the presence of the scattering by the charged impurity (nonmagnetic),
the self-energy correction and the vertex correction can be taken into account by the standard diagrammatic technique
(in first order).
We present the diagrammatical representation of the Dyson equation in Fig.7(a) and (b).
The approximated self-energy in dynamical cluster approximation is presented in Fig.7(a).
The self-energy here is the disorder-induced one (between two vertices) with the Hartree term,
which depends only on the external frequency but not on the external momentum due to the rotational invariance\cite{Goswami P}
in the presence of weak Coulomb coupling.
The electron charge density can not be viewed as constant in this case,
If the disorder is strong enough to breaks the rotational invariance, the 
nodal line in 3D Dirac or Weyl semimetal will be breaks into the nodes\cite{Moors K},
and then the momentum-dependence reappears.
It's different to the 
exchange interaction-induced one between a particle and the occupied state\cite{Altshuler B L}
which contains only the exchange interaction while the disorder-induced one contains both the exchange interaction (Coulomb)
and the impurity scattering (which yields a leading in interaction).
In fact, the disorder-induced self-energy related to the current-current correlation function while the 
Coulomb-induced self-energy related to the density-density correlation function.
These two kinds of self-energy are related by the compensation mechanism 
called Ward identity.
The relation between the reducible vertex function $\Gamma$ (complete) and the irreducible vertex function $\gamma$
can be described by the Bethe-Salpeter function
as shown in the Fig.7(c).
Comparing Fig.7(b) and (c), we know the legs in (c) is in fact the lattice Green's function.

For linear Dirac system in the disordered case, 
the momentum-independent self-energy in first-order Born approximation reads
\begin{equation} 
\begin{aligned}
\Sigma^{D}(\Omega)=\frac{1}{\hbar^{2}}\sum_{\nu}U_{k,k'}^{\nu}\int\frac{d^{2}k}{(2\pi)^{2}}\gamma_{\nu}G_{0}(k,\Omega)\gamma_{\nu'},
\end{aligned}
\end{equation}
where the summation indices $\nu=0,x,y,z$,
and the irreducible vertex function reads
\begin{equation} 
\begin{aligned}
\gamma_{\nu}=-\Omega+\hbar v_{F}{\bf k}\cdot {\pmb \sigma}+\Sigma^{D}(\Omega).
\end{aligned}
\end{equation}
The electron-impurity interaction has $U_{k,k'}=n_{i}|V_{k,k'}|^{2}$
where $n_{i}$ is the impurity concentration, and
with $V_{k,k'}$ the Fourier transform of the real space scattering potential $V_{r,r'}$ of the impurity at a random position.
The disorder potential has $V_{k,k'}(r)=V_{k,k'}\sum_{i}\delta(r-r_{i})$
and the scattering potential $V_{k,k'}$ reads $V_{k,k'}=V\sigma_{0}+{\bf I}\cdot{\pmb \sigma}$ 
in the presence of non-magnetic impurity and magnetic impurity\cite{Shiranzaei M}
where ${\bf I}=\frac{\hbar J{\bf S}}{2}$ denotes the magnetic moment
with $J$ the exchange interaction and ${\bf S}$ the spin operator of impurity.
While the Gaussian white noise distribution can be expressed by the second-order correlation
between two quenched (in equilibrium) random gauge potential
$\langle\langle V_{k,k'}(r)V_{k,k'}(r')\rangle\rangle=U_{k,k'}\delta(r-r')$.
$\gamma$ is the inreducible vertex function.
$G_{0}(k,\Omega)$ is the bare Greem's function.
The above result is consistent with the ward identity reported in Ref.\cite{Vollhardt D}
\begin{equation} 
\begin{aligned}
\Sigma(k,\Omega)=\Gamma G(k,\Omega),
\end{aligned}
\end{equation}
where the reducible vertex $\Gamma$ here has been presented in Fig.8.
In self-consistent Born approximation,
the disorder-induced self-energy would still be frequency-independent in the static limit, see \cite{Moors K}.
While for the exchange self-energies,
we integral over the frequency (in continue approximation) as
\begin{equation} 
\begin{aligned}
\Sigma^{E}(k)=\int\frac{d^{2}q}{(2\pi)^{2}}\frac{d\Omega}{2\pi}\gamma_{0} V(k,\Omega;k',\Omega')G(k-q,\Omega-\omega),
\end{aligned}
\end{equation}
where $q=k'-k$, 
$V(k,\Omega;k',\Omega')$ is the exchange-interaction potential
which can be approximated as the density-density correlation function
\begin{equation} 
\begin{aligned}
V(k,\Omega;k',\Omega')=\frac{T}{S}{\rm Tr}[\sigma_{0}G(k,\Omega)\sigma_{0}G(k',\Omega')].
\end{aligned}
\end{equation}
The noninteracting Matsubara Green's function within above self-energy reads
\begin{equation} 
\begin{aligned}
G(k,k',\Omega)=&\sum_{n,\sigma}\frac{\psi_{n}(k)\psi_{n}^{*}(k')}{\Omega+i\eta-H_{\sigma}-\Sigma^{E}(k,\Omega) }\\
=&\frac{\delta(k-k')}{\Omega+i\eta-H_{\sigma}-\Sigma^{E}(k,\Omega) },
\end{aligned}
\end{equation}
where $\delta(k-k')$ is the Kronecker $\delta$-function and will be replaced by 1 in the following.
The Landau quantitative state $\psi_{n}(k)$ satisfies
\begin{equation} 
\begin{aligned}
\psi_{n}(k)=\begin{pmatrix}
{\rm sin}\frac{\theta}{2}\Psi_{n-1}\\
{\rm cos}\frac{\theta}{2}\Psi_{n}
\end{pmatrix}
\end{aligned}
\end{equation}
with 
\begin{equation} 
\begin{aligned}
{\rm tan}\theta=-\frac{\hbar \omega_{c}\sqrt{|n|}}{D},
\end{aligned}
\end{equation}
and $\Psi_{n}$ the nomalized Harmonic oscillation function.
Here the self-energy can also be incorprated into the spectrum of Hamiltonian through the lowest-order Born approximation.

Within the lowest-order Born approximation,
the disorder-induced self-energy reads $\Sigma^{D}=U_{k,k'}/\hbar$,
and the impurity scattering rate is closely related to the DOS in Fermi level $D(E_{F})$
which vanishes when $D(E_{F})=0$.
Within this approximation
the scattering rate in Fermi surface reads
\begin{equation} 
\begin{aligned}
\tau^{-1}_{F}=&-{\rm Im}\Sigma^{D}(E_{F})\\
\simeq &-\frac{\pi}{2\hbar} D(E_{F})U_{k_{F},k_{F}'}\int\frac{d\theta}{2\pi}(1+e^{i(q_{F})\theta})^{2}(1-{\rm cos}\theta)\\
=&\frac{\pi}{2h}D(E_{F})U_{k_{F},k_{F}'}
\frac{1}{4(4q_{F}^{5}-5q_{F}^{3}+q_{F})}\\
&\left[(-8iq_{F}^{2}{\rm cos}\theta e^{i\theta q_{F}}((q_{F}^{2}-1)e^{i\theta q_{F}}+4q_{F}^{2}-1)+2i(4q_{F}^{4}-5q_{F}^{2}+1)(2i\theta q_{F}+4e^{i\theta q_{F}}+e^{2i\theta q_{F}})\right.\\
&\left.+4q_{F}{\rm sin}\theta(4q_{F}^{4}+(2-8q_{F}^{2})e^{i\theta q_{F}}-(q_{F}^{2}-1)e^{2i\theta q_{F}}-5q_{F}^{2}+1))\right]\bigg|_{\theta},
\end{aligned}
\end{equation}
which implies that the case of backscattering ($\theta=\pi$ and $k=-k'$) and forward scattering ($\theta=0$) can't happen
since the scattering momentum $q_{F}=|k_{F}-k_{F}'|=2|k_{F}|{\rm sin}(\theta/2)$.
That reveals that the scattering rate is contributed by the field renormalization 
in the disordered configuration
but unaffected by the Dirac-mass renormalization which emerges in the real part of the self-energy\cite{Goswami P}.
Base on a reasonable setting of the parameters,
the above relaxation rate is in an order of $10^{12}s^{-1}$\cite{Wu C HElectronic}.

Next we discuss the Green's function in real space with the self-energy effect.
For gapped parabolic system at half-filling, the disorder-induced self-energy can be obtained as
\begin{equation} 
\begin{aligned}
\Sigma^{D}(\Omega)=\frac{U_{k,k'}\pi}{\hbar^{2}4\pi^{2}}(\frac{f}{\sqrt{D^{2}-(\Omega+i\eta)^{2}}}),
\end{aligned}
\end{equation}
where 
\begin{equation} 
\begin{aligned}
f=&-m\left[ 
(\Omega+i\eta){\rm tan}^{-1}(\frac{k^{2}}{2m\sqrt{D^{2}-(\Omega+i\eta)^{2}}})\right.\\
&\left.+\sqrt{D^{2}-(\Omega+i\eta)^{2}}{\rm ln}(\sqrt{4D^{2}m^{2}+k^{4}}+k^{2})\right.\\
&\left.+(\Omega+i\eta){\rm tan}^{-1}(\frac{k^{2}(\Omega+i\eta)}{\sqrt{4D^{2}m^{2}+k^{4}}\sqrt{D^{2}-(\Omega+i\eta)^{2}}}).
\right],
\end{aligned}
\end{equation}
For gapless parabolic Dirac system (in metal phase),
we have
\begin{equation} 
\begin{aligned}
\Sigma^{D}(\Omega)=\frac{U_{k,k'}\pi}{\hbar^{2}4\pi^{2}}(-m{\rm ln}(\Lambda^{2}-2m(\Omega+i\eta)))
+\frac{U_{k,k'}\pi}{\hbar^{2}4\pi^{2}}(m{\rm ln}(2m(\Omega+i\eta))),
\end{aligned}
\end{equation}
where the results are presented in the Fig.9.
In Fig.9(a),
we present the diagrammatic representation of the self-energy correction,
in Fig.9(b), we show the disorder-induced self-energy as a function of the frequency $\Omega$ and energy-cutoff $\Lambda$,
and Fig.9(c) corrsponds to the case of $\Lambda=4.8$ eV.
Then the real space Green's function (with self-energy effect) reads
\begin{equation} 
\begin{aligned}
G(\pm R,\Omega)=&\int\frac{d^{2}k}{(2\pi)^{2}}e^{\pm i{\bf k}\cdot{\rm R}}G(k,\Omega)\\
=&\pi I_{0}(\pm ikR)\int^{\Lambda}_{0}\frac{k}{\Omega+i\eta-\frac{k^{2}}{2m}-\Sigma^{D}(\Omega)},
\end{aligned}
\end{equation}
where $\pm R=|r-r'|(|r'-r|)$ and $I_{0}$ is the modified Bessel function of the first kind,
the above integral is can be solved approximatedly as
\begin{equation} 
\begin{aligned}
G(\pm R,\Omega)
\approx &
\frac{\pi m}{4}[R^{2}(k^{2}-2m(i\eta-\Sigma^{D}(\Omega)+\Omega))\\
&+2(-2+mR^{2}(i\eta-\Sigma^{D}(\Omega)+\Omega)){\rm ln}(k^{2}-2m(i\eta-\Sigma^{D}(\Omega)+\Omega))]\bigg|^{\Lambda}_{0},
\end{aligned}
\end{equation}
where we use the series expansion of the $I_{0}(\pm ikR)$ near $R=0$.
Another way to solve above integral is by using the Rayleigh equation\cite{Hosseini M V}
\begin{equation} 
\begin{aligned}
e^{ i{\bf k}\cdot{\rm R}}=4\pi\sum^{\infty}_{l=0}\sum^{l}_{m=-l}i^{l}j_{l}(kR)Y^{*}_{lm}(\theta_{R},\phi_{R})Y_{lm}(\theta_{k},\phi_{k}),
\end{aligned}
\end{equation}
where $j_{l}(kR)$ is the spherical Bessel function,
$Y_{lm}$ is the spherical harmonic function:
\begin{equation} 
\begin{aligned}
Y_{lm}(\theta_{R},\phi_{R})=[A_{lm}(R){\rm cos}m\phi_{R}+B_{lm}(R){\rm sin}m\phi_{R}]P^{m}_{l}({\rm cos}\theta_{R}),
\end{aligned}
\end{equation}
with $P^{m}_{l}$ the associated Legendre function.
Through Eq.(36), we can know that the real space Green's function of the gapless parabolic Dirac system in isotropic case
(with Berry phase $2\pi$) does not contains the Pauli matrix ${\pmb \sigma}$ unlike the linear Dirac system (monolayer)\cite{Wu C Hrkky,
Wu C Hcurrent,Zare M}
or the 3D Dirac/Weyl semimetal\cite{Chang H R}.
For the gapless linear Dirac system (with Berry phase $\pi$),
the self-energy reads
\begin{equation} 
\begin{aligned}
\Sigma^{D}(\Omega)=&\frac{U_{k,k'}\pi}{\hbar^{2}4\pi^{2}}\int^{\Lambda}_{0}kdk\frac{1}{\Omega+i\eta-\hbar v_{F}{\bf k}\cdot{\pmb \sigma}}\\
=&\frac{U_{k,k'}\pi}{\hbar^{2}4\pi^{2}}\int^{\Lambda}_{0}kdk\frac{(\Omega+i\eta)\pm \hbar v_{F}{\bf k}\cdot{\pmb \sigma}}
{(\Omega+i\eta)^{2}-\hbar^{2} v^{2}_{F} k^{2}}\\
=&\frac{U_{k,k'}\pi}{\hbar^{2}4\pi^{2}}
[-\frac{(\Omega+i\eta){\rm ln}(-\hbar^{2}v_{F}^{2}k^{2}+(\Omega+i\eta)^{2})}{2\hbar^{2}v_{F}^{2}}]\bigg|_{0}^{\Lambda}\\
&\pm
{\pmb \sigma}\cdot{\bf e}_{R}\frac{U_{k,k'}\pi}{\hbar^{2}4\pi^{2}}
[-\frac{{\rm cos}\theta{\rm ln}(-\hbar^{2}v_{F}^{2}k^{2}+(\Omega+i\eta)^{2})}{2\hbar v_{F}}]\bigg|_{0}^{\Lambda},
\end{aligned}
\end{equation}
where $\theta$ is the angle between vectors $k$ and $R$,
and ${\bf e}_{R}$ denotes the direction of $R$.
By substituting the above self-energy into the equation
\begin{equation} 
\begin{aligned}
G(\pm R,\Omega)=&\int\frac{d^{2}k}{(2\pi)^{2}}e^{\pm i{\bf k}\cdot{\rm R}}G(k,\Omega)\\
=&\pi I_{0}(\pm ikR)\int^{\Lambda}_{0}\frac{k}{\Omega+i\eta-\hbar v_{F}{\bf k}\cdot{\pmb \sigma}-\Sigma^{D}(\Omega)},
\end{aligned}
\end{equation}
we can then obtain the real space Green's function of the linear Dirac system.
Different to the linear response theory where the Dzyaloshinskii-Moriya interaction vanishes,
the nonlinear response in topological insulator with nonzero SOC can leads to the nonlinear spin susceptibility and
the nonlinear RKKY interaction in the presence of the broken inversion symmetry.
We can also extend to the multilayer Dirac system with linear relation (semimetal),
base on the assumation of the isotropic dispersion,
the RKKY Range function at finite temperature and static limitreads\cite{Park S}
\begin{equation} 
\begin{aligned}
\Pi_{RKKY}(R,T)=\int^{k_{c}}_{0}\frac{1}{2\pi^{2}}q^{2}j_{0}(qR)\Pi(q,T)dq,
\end{aligned}
\end{equation}
where $k_{c}$ is the momentum cutoff which related to the $\Lambda$.
$\Pi(q,T)$ is the static polarization at finite temperature (here we only consider the longitudinal spin susceptibility),
which reads
\begin{equation} 
\begin{aligned}
\Pi(q,T)=\frac{\Pi(q,0)}{\mathcal{T}},
\end{aligned}
\end{equation}
where 
\begin{equation} 
\begin{aligned}
\Pi(q,0)=\frac{-q^{2}}{24\pi^{2}\hbar v_{F}}{\rm ln}\frac{4k_{c}^{2}}{q^{2}}
\end{aligned}
\end{equation}
in the case of $\mu=0$ and $D=0$,
and the polarization function here is proportional to the DOS and has $\Pi(q,0)\propto q^{2}$
which is consistent with the result of Ref.\cite{Park S},
and we can also imaging that for parabolic 3D Dirac system it has $\Pi(q,0)\propto q$.
$\mathcal{T}$ is the temperature-dependent factor, which reads
\begin{equation} 
\begin{aligned}
\mathcal{T}=4T{\rm cosh}^{2}(\frac{E_{F}(T)}{2T}),
\end{aligned}
\end{equation}
here the temperature-dependent Fermi energy is $E_{F}(T)=T{\rm ln}[e^{\frac{n}{D_{F}^{(3)}T}}-1]$\cite{Liu Y}
where $n$ is the carrier density,
and the three-dimensional DOS at Fermi level can be approximated as $D_{F}^{(3)}=\sqrt[3]{\frac{9g_{s}g{v}n^{2}}{2\pi^{2}}}\frac{1}{v_{F}}$.
After some algebra,
the RKKY range function can be obtained as
\begin{equation} 
\begin{aligned}
\Pi_{RKKY}(R,T)=&
-\frac{1}{48\pi^{4}\hbar R^{5}\mathcal{T}v_{F}}
[{\rm sin}(qR)(3(q^{2}R^{2}-2){\rm ln}(\frac{4k_{c}^{2}}{q^{2}})-2q^{2}R^{2}+22)\\
&-qR{\rm cos}(qR)((q^{2}R^{2}-6){\rm ln}(\frac{4k_{c}^{2}}{q^{2}})+10)
-12{\rm Si}(qR)]
\bigg|_{0}^{k_{c}}
\end{aligned}
\end{equation}
where ${\rm Si}(z)$ denotes the sine integral.
The above expression approximately provides the long range ($R\gg a$) decay of the RKKY range function
for the 3D (multilayer) gapless Dirac system, which is $\sim\frac{{\rm sin}(qR)}{R^{3}}$
where the scattering momentum $q$ here is usually at the Fermi surface (i.e., $q=2k_{F}$) where the scattering the dominant,
and it's also similar to the results about the Friedel oscillation of the screened potential
\cite{Wu C HDynamical,Wu C HInterband,Chang H R,Wu C Hcurrent,Wu C HRINP}
as well as the long range spin susceptibility\cite{Stano P}.
In Fig.10,
we show the RKKY range function for the casees of $\mathcal{T}=1$ and $\mathcal{T}=2$,
where we set the momentum cutoff as $k_{c}=4.8$ eV.
We can see that the fluctuation is smaller at higher temperature,
and since the oscillation period $A$ is related to the scattering mometum by $A=2\pi/q$,
the smaller the $q$ is, the slower the fluctuation of the range function.
It's also proved\cite{Park S} that,
once the distance exceeds the mean-free path $\ell$,
the effective RKKY interaction decays as $\sim e^{-(R-\ell)/\ell}$.

\section{Spectral function in the presence of self-energy effect}

The spectral function in the presence of relaxation-induced broaden can be obtained as
\begin{equation} 
\begin{aligned}
A(\Omega)=-\frac{1}{\pi}{\rm Im}\frac{1}{\Omega+i\eta-\varepsilon-\Sigma^{D}(\Omega)}.
\end{aligned}
\end{equation}
We present in Fig.11 the results of the
spectral function in (single-particle) first-order Born approximation.
We can see that for a fixed momentum $k$,
the spectral function has only one peak due to the single particle approximation,
and the trajectory of the positions of teh peaks is parabolic with $k$ for the parabolic Dirac system,
and such trajectory will becomes linear with $k$ for the linear Dirac system (e.g.,
see Ref.\cite{Ando T}).
That also implies that the spectral function is related to not only the quasiparticle relaxation (scattering),
but also to the band dispersion.
Besides, since the nonequilibrium many-body Green's function is hard to solved in a realistic system\cite{Stauber T},
we usually use the equilibrium single-particle Green's function as we done above (i.e., 
the self-consistent first-order Born approximation),
however, sometimes the overlap between the noninteracting and interating states 
(in different Fermion subspaces) needed to be taken into account to considering the effects of freqeuncy including the Rabi oscillation
as well as the detuning of the $\delta$-function.
The related approximation is the rotating wave approximation when the frequency is close to the resonance one,
and asymptotic rotating wave approximation\cite{Kumar V} when the frequency is far away from the resonance,
e.g., when the frequency is larger than 1200 THz (corresponds to the electron-hole energy 4.8 eV) for silicene
\cite{Wu C HGeometrical,Wu C HTight,Wu C HAnomalous}.
Although the multipeak feature emerges in the exact spectral function (contains the broadened $\delta$-term),
e.g., like the one calculated within the rotating wave approximation\cite{Stauber T},
the parabolic trajectory of the position of the main peak remains,
which as a intrinsic difference between the linear Dirac system.
The spetral function is related to the retarded two-point Green's function by
\begin{equation} 
\begin{aligned}
iG^{R}_{ij}(t-t')=\theta(t-t')[iG^{>}_{ij}(t-t')-iG^{<}_{ij}(t-t')],
\end{aligned}
\end{equation}
where 
\begin{equation} 
\begin{aligned}
iG^{>}_{ij}(t-t')=&\frac{-i}{2\pi}\int e^{-i\Omega t}2\pi iA(\Omega)[1-N_{F}(\Omega)]\\
=&\langle c_{i}(t)c_{j}^{\dag}(t')\rangle,\\
-iG^{<}_{ij}(t-t')=&\frac{-i}{2\pi}\int e^{-i\Omega t}2\pi iA(\Omega)N_{F}(\Omega)\\
=&\langle c^{\dag}_{j}(t')c_{i}(t)\rangle.
\end{aligned}
\end{equation}
While for the Coulomb interaction-induced self-energy in Hartree-Fock approximation
which is frequency-independent,
and in first-order perturbation theory reads\cite{Sodemann I,Vafek O}
\begin{equation} 
\begin{aligned}
\Sigma^{E}(k)=\frac{e^{2}{\bf k}\cdot{\pmb \sigma}}{4\epsilon}{\rm ln}\frac{\Lambda}{k},
\end{aligned}
\end{equation}
where $\epsilon$ is the dielectric constant of the environment.
Such first-order self-energy correction is diagrammatically presented in the first term of the Fig.9(a).
Such self-energy correction logarithmically enhance the Dirac velocity as
$v_{k}=v_{F}+\frac{e^{2}}{4\epsilon}{\rm ln}\frac{\Lambda}{k}$,
and it's persistent to the second order expansion of the Coulomb interaction (or the $U_{k,k'}$).

\section{Summary}

In summary, we investigate the 
electronic properties of the parabolic Dirac system in the presence of impurity scattering,
where the non-adiabatic correction (Berry curvature) and the self-energy correction are considered.
Our discussions are not confined to the zero-temperature case,
and are meaningful to the exploration of the electronic properties in Dirac or Weyl systems.

\end{large}
\renewcommand\refname{References}

\clearpage

Fig.1
\begin{figure}[!ht]
   \centering
 \centering
   \begin{center}
     \includegraphics*[width=1\linewidth]{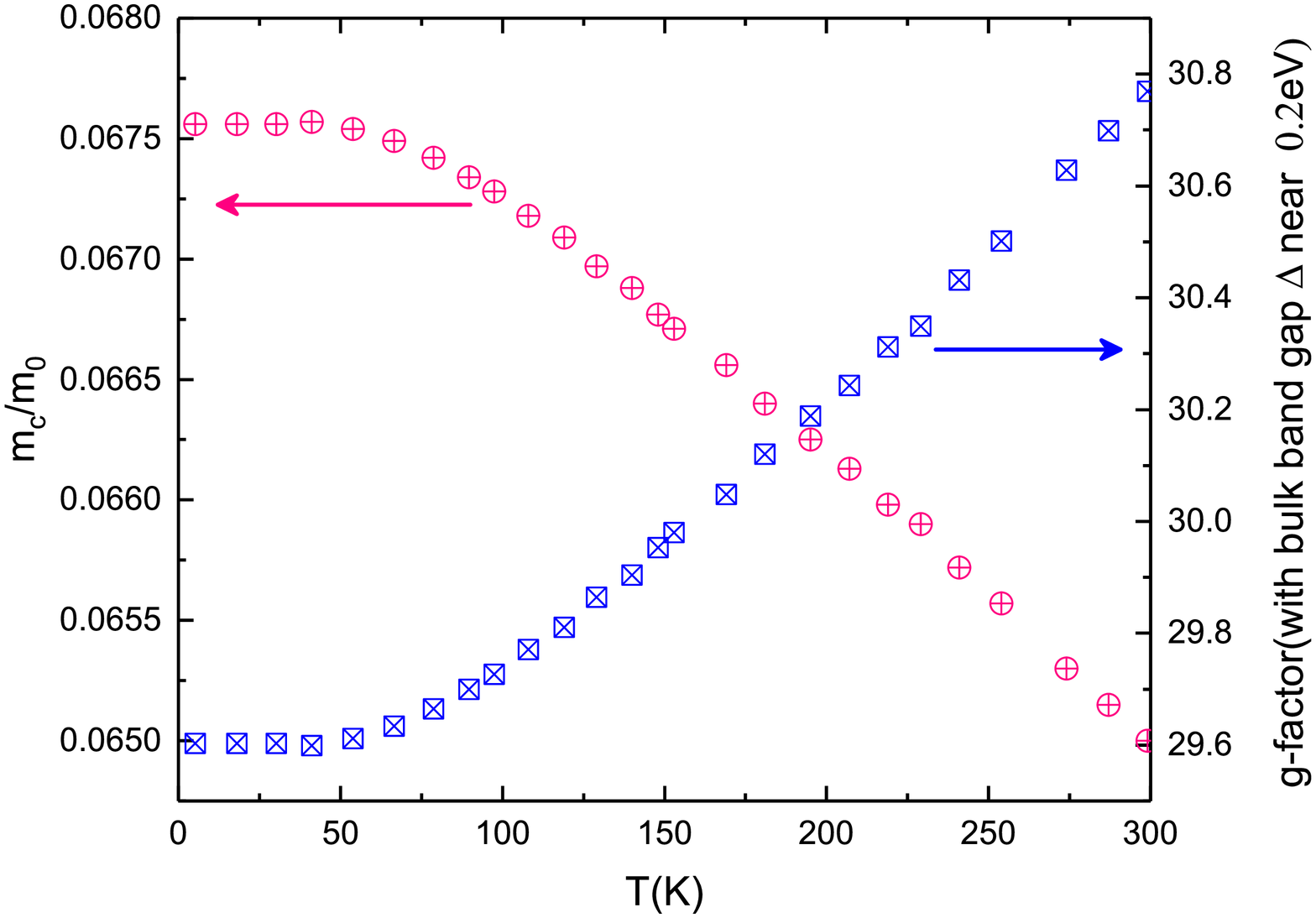}
\caption{(Color online) Effective mass $m_{c}$ and the $g$-factor in three-dimensional electron system GaAs\cite{Hubner J}.
}
   \end{center}
\end{figure}
\clearpage
Fig.2
\begin{figure}[!ht]
   \centering
 \centering
   \begin{center}
     \includegraphics*[width=1\linewidth]{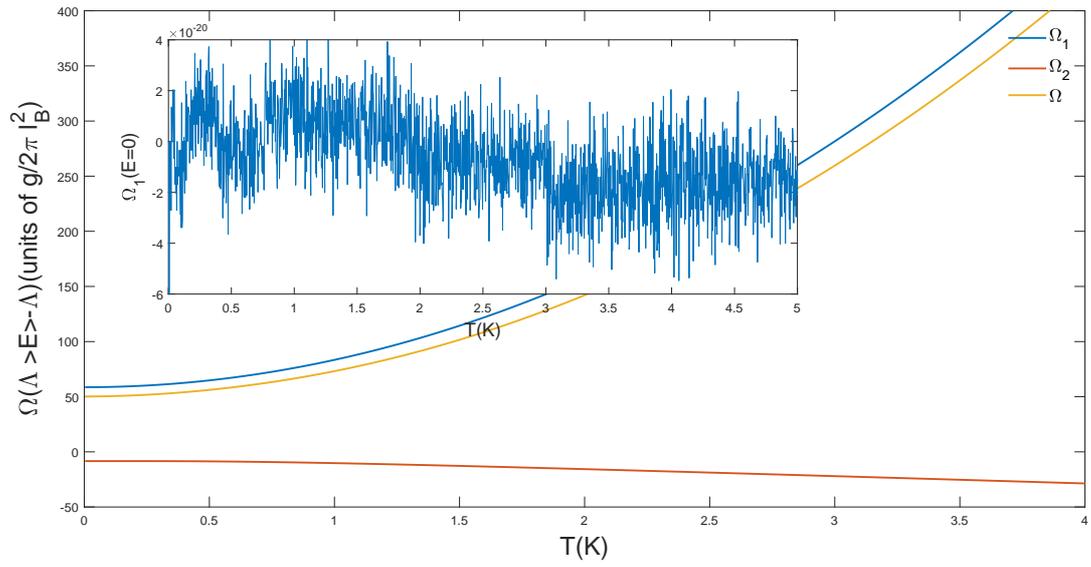}
\caption{(Color online) Thermodynamical potential as a function of temperature.
The inset shows the fluctuation of $\Omega_{1}$ in zero energy case.
}
   \end{center}
\end{figure}
\clearpage

Fig.3
\begin{figure}[!ht]
   \centering
 \centering
   \begin{center}
     \includegraphics*[width=0.6\linewidth]{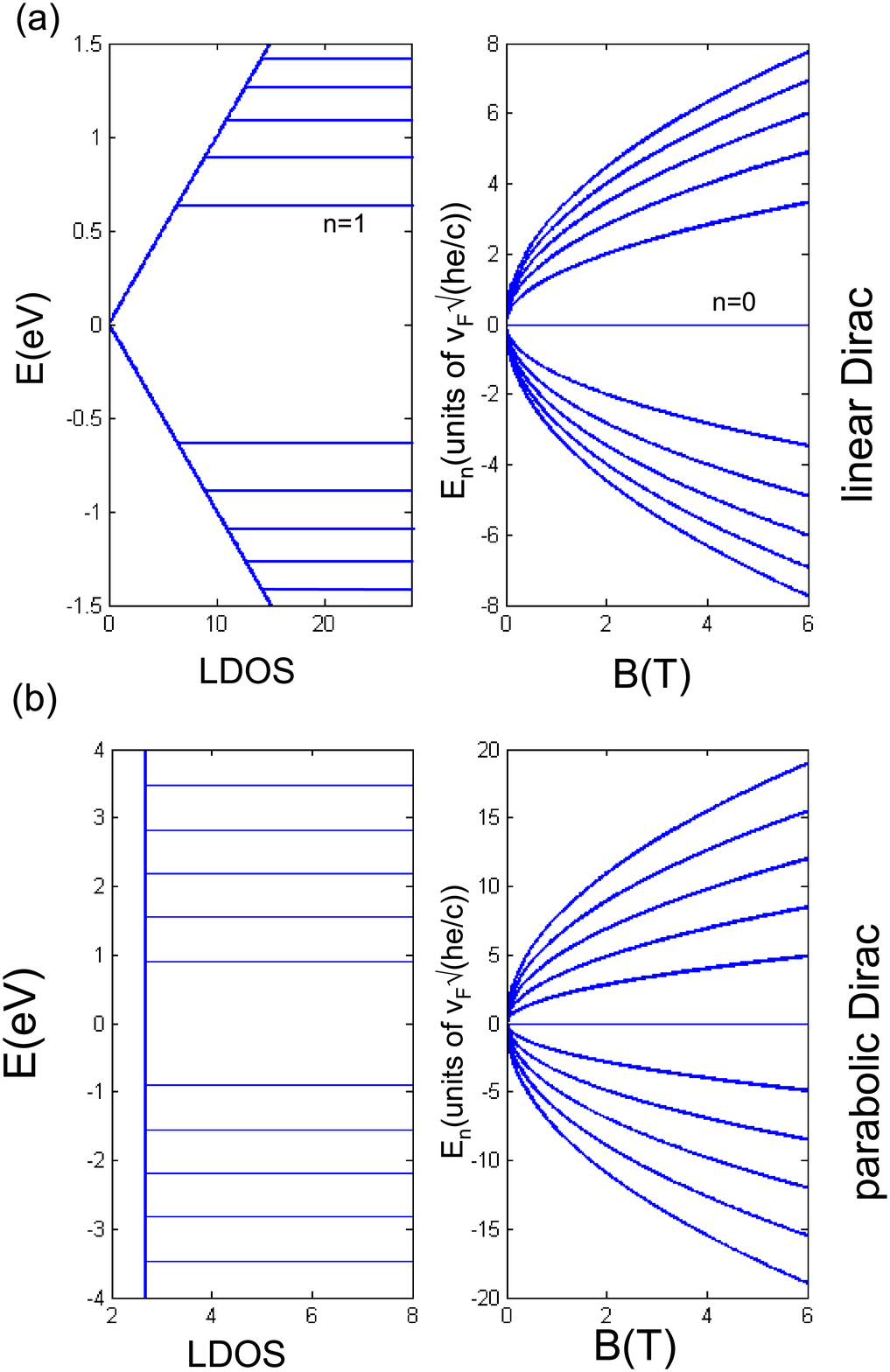}
\caption{(Color online) Local DOS (LDOS) as a function of energy $E$ (left panel) and LLs as a function of $B$ (right panel)
of the linear Dirac system (a) and parabolic Dirac system (b).
}
   \end{center}
\end{figure}
\clearpage

Fig.4
\begin{figure}[!ht]
   \centering
 \centering
   \begin{center}
     \includegraphics*[width=0.6\linewidth]{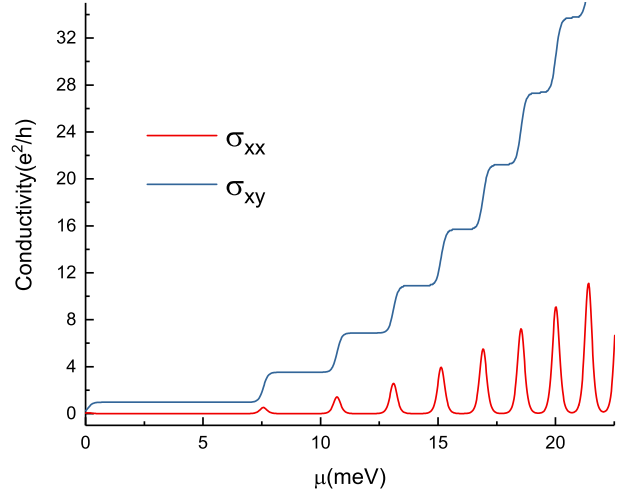}
\caption{(Color online) Hall conductivity $\sigma_{xy}$ and the longitudinal conductivity $\sigma_{xx}$ as a function of Fermi energy.
The temperature setted as $T=0.1$ K, and the magnetic field setted as $B=1$ T.
}
   \end{center}
\end{figure}

Fig.5
\begin{figure}[!ht]
   \centering
 \centering
   \begin{center}
     \includegraphics*[width=0.6\linewidth]{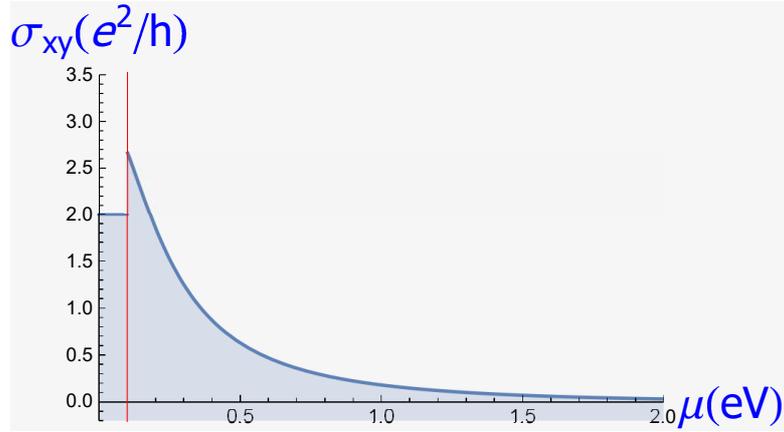}
\caption{(Color online) The intrinsic Hall conductivity of the bilayer Dirac system contributed by the Berry curvature.
The red line indicates the value of Dirac-mass,
the Hall conductivity dependents only on natural
constant and the sign of Dirac-mass for $\mu<D$,
and it's $2e^{2}/h$ which is twice of that of the monolayer 2D Dirac system.
}
   \end{center}
\end{figure}
\clearpage

Fig.6
\begin{figure}[!ht]
   \centering
 \centering
   \begin{center}
     \includegraphics*[width=0.6\linewidth]{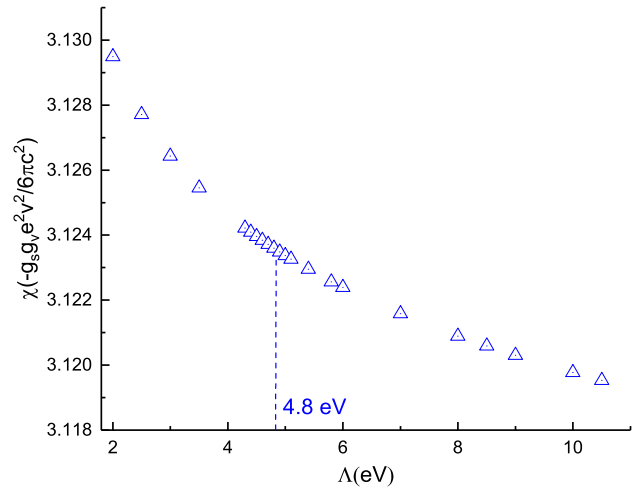}
\caption{(Color online) Orbital diamagnetic susceptibility under the short-range disorder as a function of the energy cutoff $\Lambda$.
}
   \end{center}
\end{figure}
\clearpage

Fig.7
\begin{figure}[!ht]
   \centering
 \centering
   \begin{center}
     \includegraphics*[width=0.6\linewidth]{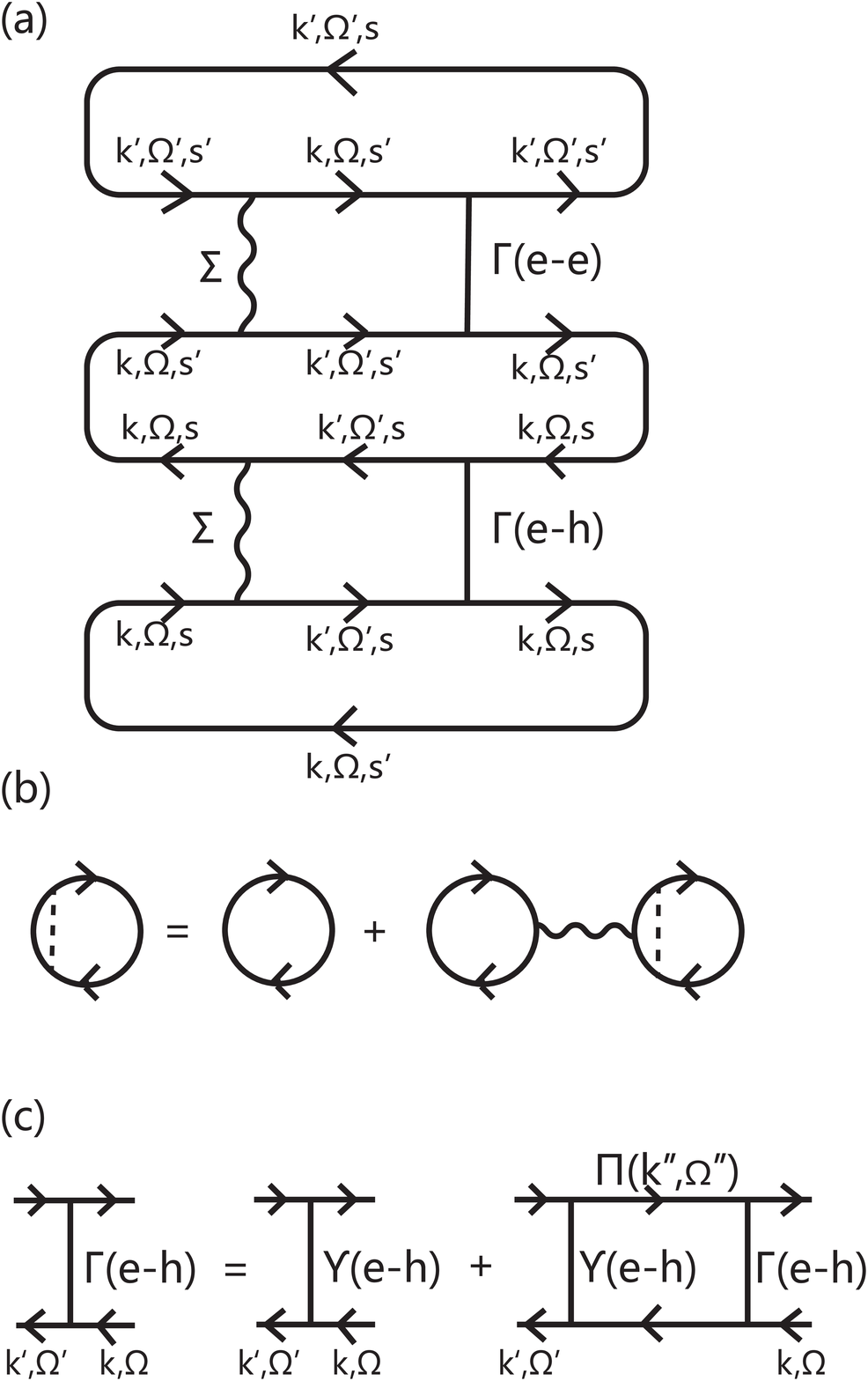}
\caption{(Color online) Diagrammatical representation of the Dyson equation ((a) and (b)) with approximated self-energy (on the large cluster)
$\Sigma$ (the wavy line)
and the vertex function (the straight line) $\Gamma$ (reducible) and $\gamma$ (inreducible).
$k'=k+q$ is the momentum after scattering, $\Omega'=\Omega+\omega$ is the frequency (by using the 
retarded form analytical continuation) after scattering,
where $\Omega=(2n+1)\pi T$ is the Fermion Matsubara frequency.
$s$ is the transverse spin.
The e-h and e-e denote the electron-hole vertex function and electron-electron vertex function, respectively.
(b) describes the Dyson equation
$G=G_{0}+G_{0}*\Sigma^{D}*G=G_{0}+G*\Sigma^{D}*G_{0}$\cite{Wu C Htime}.
In (b), 
the solid circle represents the disorder-averaged retarded Green function and the dash line within the circle denotes the Coulomb interation.
Note that here the wavy line between the bare Green's function and the 
 disorder-averaged retarded Green function is not the dynamical screened Coulomb interaction but the disorder-induced
Hartree self-energy correction in the presence of impurity scattering.
(c) is the Bethe-Salpeter equation which describes the relation between the reducible vertex function and inreducible vertex function,
and $\Pi(k'',\Omega)$ is the bare one-loop polarization function (i.e., the density-density correlation function here
and describes the reducible part of the $\Gamma$).
}
   \end{center}
\end{figure}
\clearpage

Fig.8
\begin{figure}[!ht]
   \centering
 \centering
   \begin{center}
     \includegraphics*[width=0.6\linewidth]{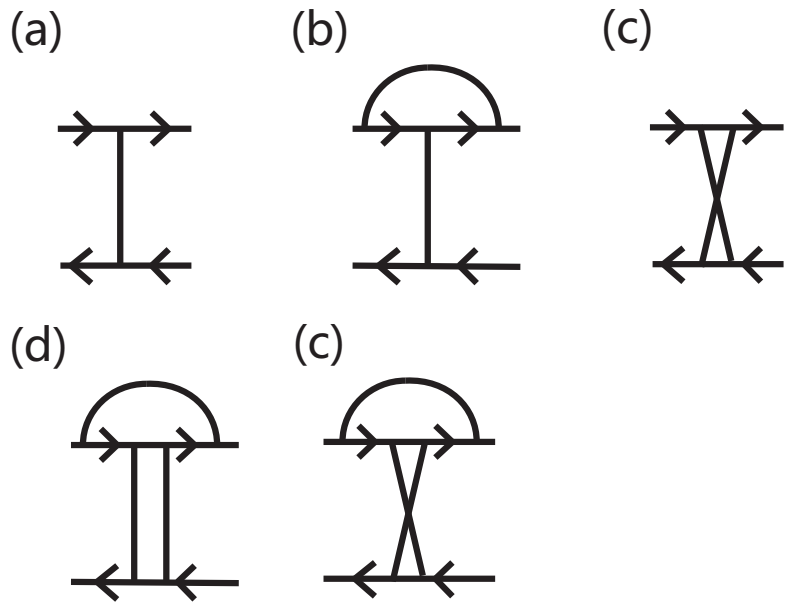}
\caption{(Color online) First-order (a), second-order (b,c), and third-order (c,d) component of the vertex function in Eq.(26).
}
   \end{center}
\end{figure}
\clearpage

Fig.9
\begin{figure}[!ht]
\subfigure{
\begin{minipage}[t]{0.5\textwidth}
\centering
\includegraphics[width=1\linewidth]{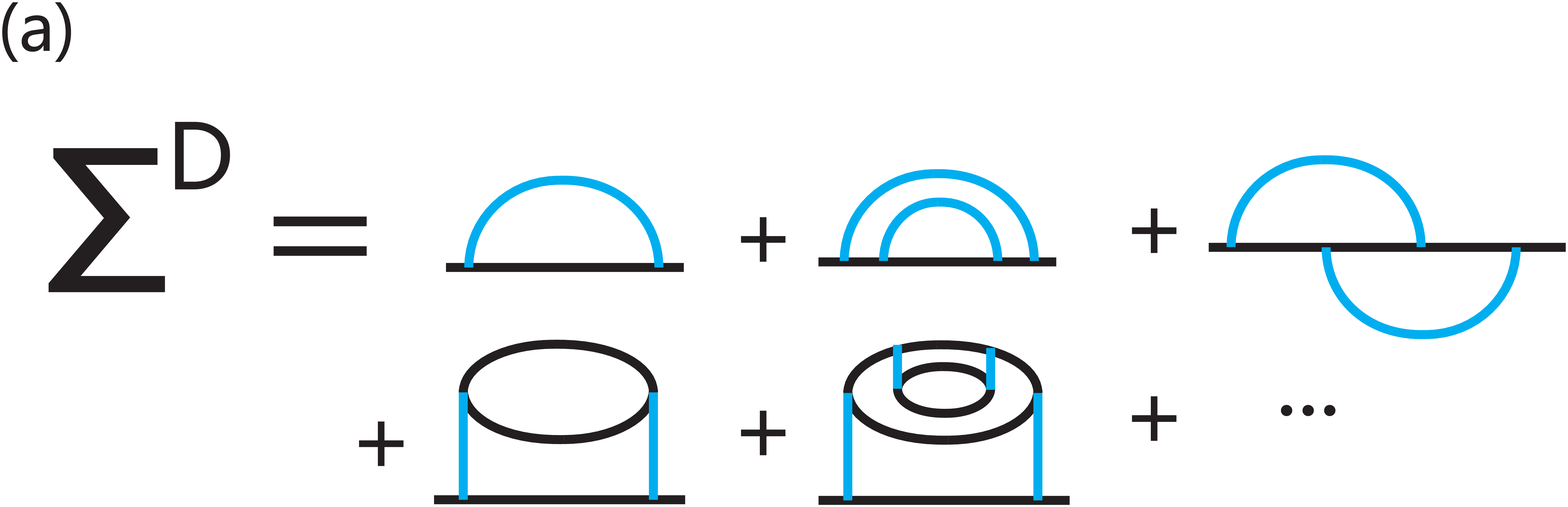}
\label{fig:side:a}
\end{minipage}
}\\
\subfigure{
\begin{minipage}[t]{0.55\textwidth}
\centering
\includegraphics[width=0.9\linewidth]{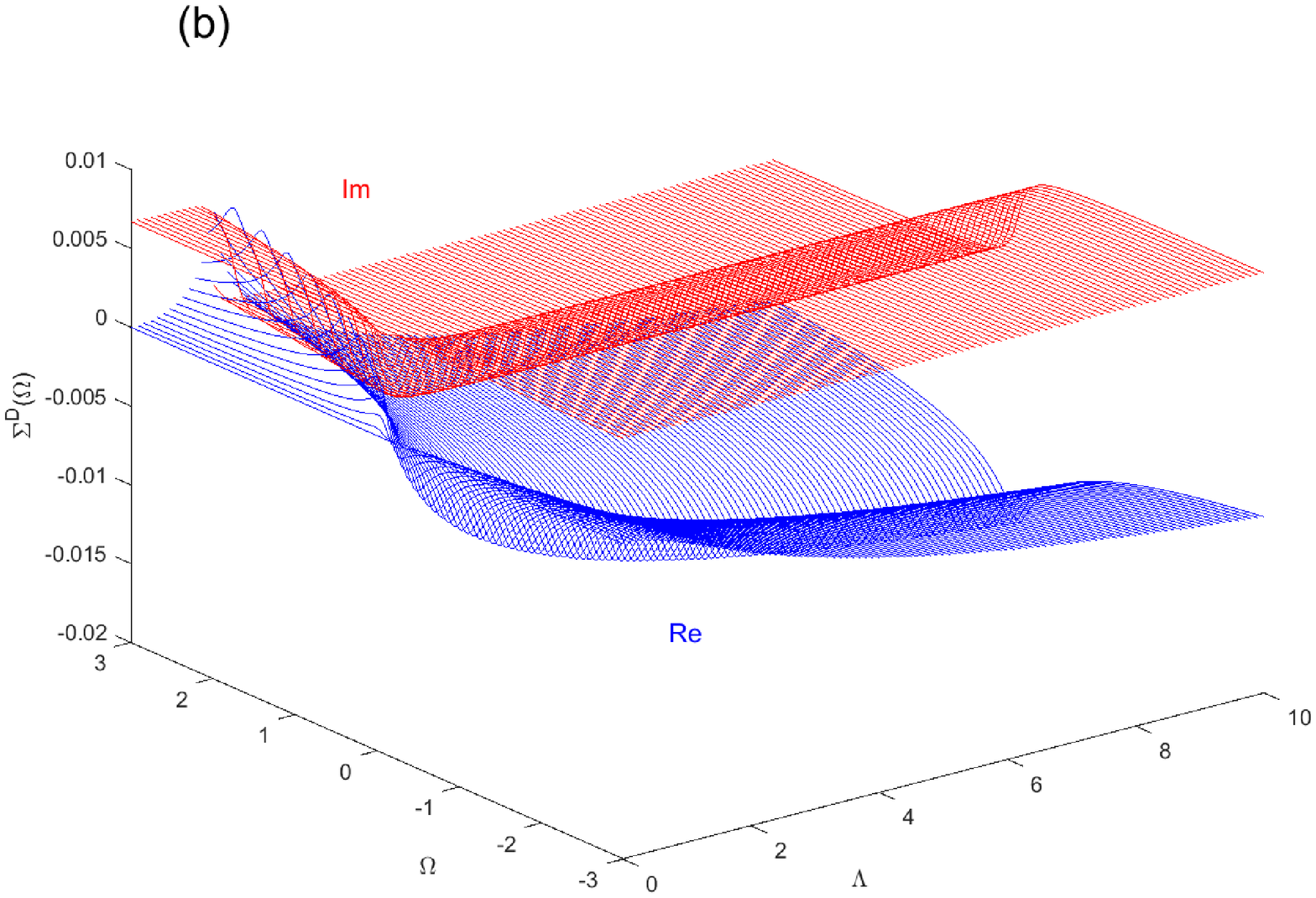}
\label{fig:side:b}
\end{minipage}
}\\
\subfigure{
\begin{minipage}[t]{0.55\textwidth}
\centering
\includegraphics[width=0.9\linewidth]{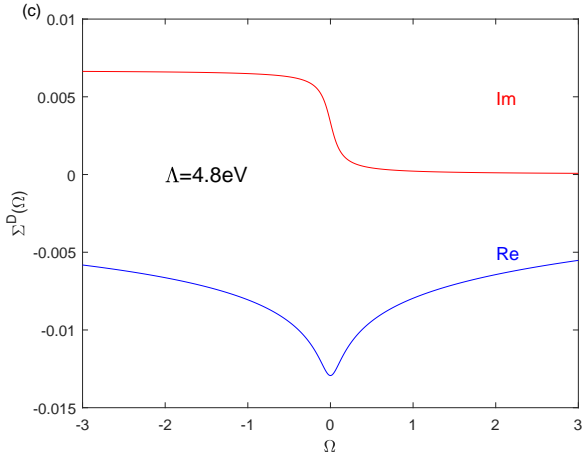}
\label{fig:side:b}
\end{minipage}
}
\caption{(Color online) 
(a) Diagrammatic representation of the self-energy
which is devided into five terms.
We note that the second term is related to the impurity-disorder,
and the foyrth term is related to the RPA\cite{Mishchenko E G}.
The blue lines can be replaced by the dash-line or the wavy-line which corresponds to the disorder-induced and exchange-induced self-energy,
respectively.
(b) is the disorder-induced self-energy in zero-temperature limit
as a function of the frequency and energy cutoff.
(c) shows the self-energy at cutoff $\Lambda=4.8$ eV which is nearly twice of the $\pi$-band width
of the 2D Dirac systems.
}
\end{figure}
\clearpage

Fig.10
\begin{figure}[!ht]
\subfigure{
\begin{minipage}[t]{0.5\textwidth}
\centering
\includegraphics[width=1\linewidth]{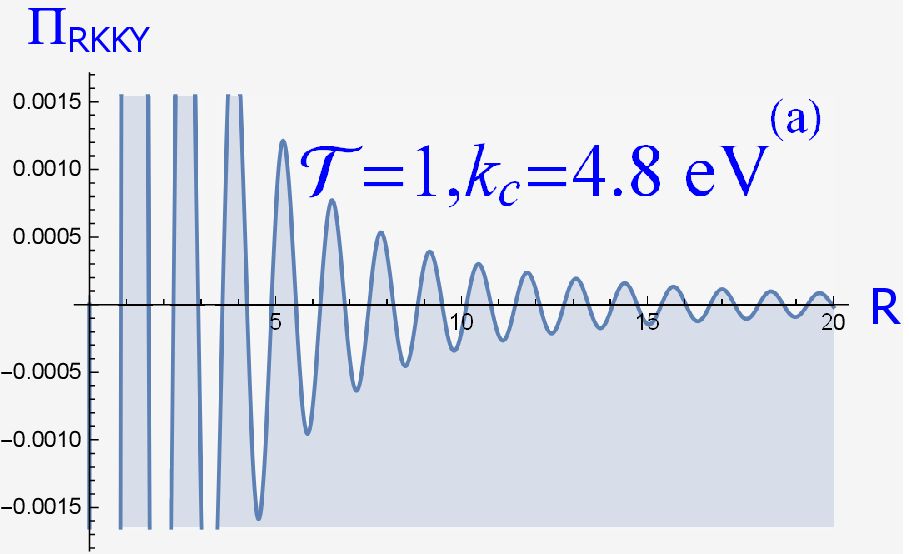}
\label{fig:side:a}
\end{minipage}
}\\
\subfigure{
\begin{minipage}[t]{0.55\textwidth}
\centering
\includegraphics[width=0.9\linewidth]{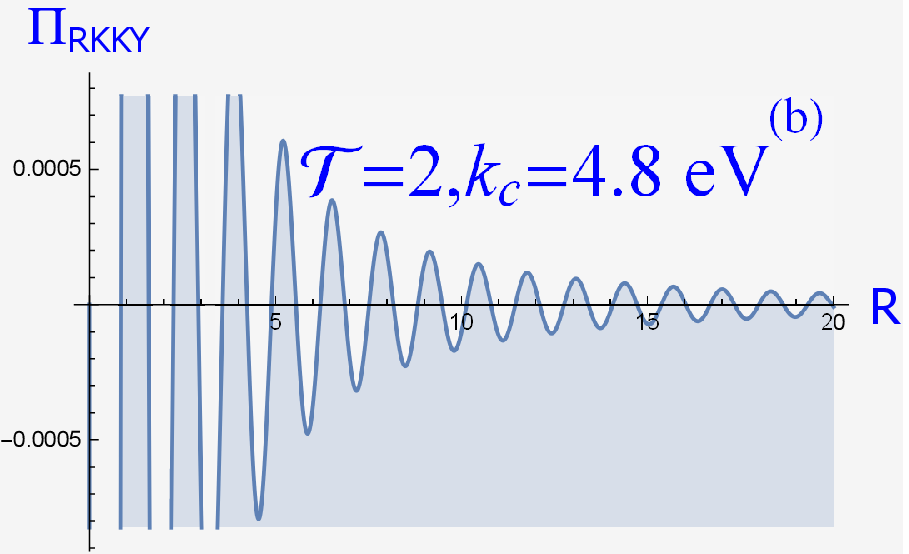}
\label{fig:side:b}
\end{minipage}
}\\
\caption{(Color online) The RKKY range function with $\mathcal{T}=1$ (a) and $\mathcal{T}=2$ (b).
}
\end{figure}
\clearpage

Fig.11
\begin{figure}[!ht]
   \centering
 \centering
   \begin{center}
     \includegraphics*[width=0.6\linewidth]{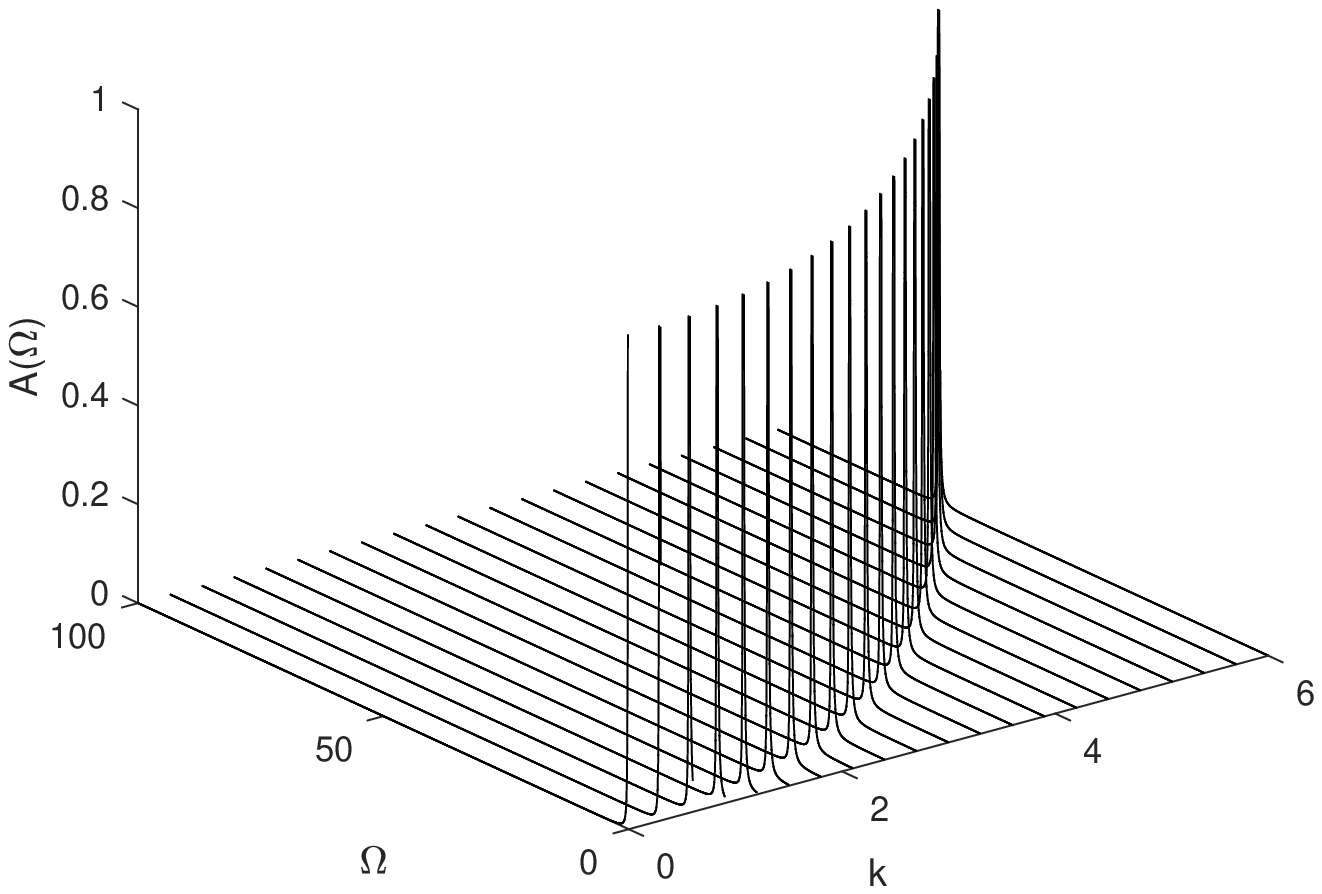}
\caption{(Color online) Spectral function in self-consistent first-order Born approximation
as a function of frequency $\Omega$ and momentum $k$.
Here the broadening is setted as $\eta=0.1\hbar\omega_{c}$ and $U_{k,k'}=0.1$ eV.
}
   \end{center}
\end{figure}


\begin{thebibliography}{99}

\bibitem{McCann E}McCann E, Fal’ko V I. Landau-level degeneracy and quantum Hall effect in a graphite bilayer[J]. Physical Review Letters, 2006, 96(8): 086805.

\bibitem{Chadov S}Chadov S, Qi X, K{\"u}bler J, et al. Tunable multifunctional topological insulators in ternary Heusler compounds[J]. Nature materials, 2010, 9(7): 541.
\bibitem{Wan W}Wan W, Ge Y, Yang F, et al. Phonon-mediated superconductivity in silicene predicted by first-principles density functional calculations[J]. EPL (Europhysics Letters), 2013, 104(3): 36001.
\bibitem{Teshome T}Teshome T, Datta A. Topological Insulator in Two-Dimensional SiGe Induced by Biaxial Tensile Strain[J]. ACS Omega, 2018, 3(1): 1-7.
\bibitem{Teshome T2}Teshome T, Datta A. Phase Coexistence and Strain-Induced Topological Insulator in Two-Dimensional BiAs[J]. The Journal of Physical Chemistry C, 2018.
\bibitem{Yu R}Yu R, Qi X L, Bernevig A, et al. Equivalent expression of Z 2 topological invariant for band insulators using the non-abelian Berry connection[J]. Physical Review B, 2011, 84(7): 075119.

\bibitem{Li X}Li X, Zhang Z, Yao Y, et al. High throughput screening for two-dimensional topological insulators[J]. 2D Materials, 2018, 5(4): 045023.
\bibitem{Haratipour N}Haratipour N, Liu Y, Wu R J, et al. Mobility Anisotropy in Black Phosphorus MOSFETs With HfO₂ Gate Dielectrics[J]. IEEE Transactions on Electron Devices, 2018 (99): 1-9.


\bibitem{Ahn S}Ahn S, Hwang E H, Min H. Collective modes in multi-Weyl semimetals[J]. Scientific reports, 2016, 6: 34023.
\bibitem{Ghosal A}Ghosal A, Goswami P, Chakravarty S. Diamagnetism of nodal fermions[J]. Physical Review B, 2007, 75(11): 115123.
\bibitem{Shakouri K}Shakouri K, Vasilopoulos P, Vargiamidis V, et al. Integer and half-integer quantum Hall effect in silicene: Influence of an external electric field and impurities[J]. Physical Review B, 2014, 90(23): 235423.
\bibitem{Park J}Park J, Pasupathy A N, Goldsmith J I, et al. Coulomb blockade and the Kondo effect in single-atom transistors[J]. Nature, 2002, 417(6890): 722.
\bibitem{Yang K}Yang K. Spontaneous symmetry breaking and quantum Hall effect in graphene[J]. Solid State Communications, 2007, 143(1-2): 27-32.
\bibitem{Fuseya Y}Fuseya Y, Ogata M, Fukuyama H. Transport properties and diamagnetism of dirac electrons in bismuth[J]. Journal of the Physical Society of Japan, 2014, 84(1): 012001.
\bibitem{Hubner J}H{\"u}bner J, D{\"o}hrmann S, H{\"a}gele D, et al. Temperature-dependent electron Landé g factor and the interband matrix element of GaAs[J]. Physical Review B, 2009, 79(19): 193307.
\bibitem{Orlita M}Orlita M, Basko D M, Zholudev M S, et al. Observation of three-dimensional massless Kane fermions in a zinc-blende crystal[J]. Nature Physics, 2014, 10(3): 233.
\bibitem{Koshino M}Koshino M, Ando T. Anomalous orbital magnetism in Dirac-electron systems: Role of pseudospin paramagnetism[J]. Physical Review B, 2010, 81(19): 195431.
\bibitem{Schafgans A A}Schafgans A A, Post K W, Taskin A A, et al. LL spectroscopy of surface states in the topological insulator Bi 0.91 Sb 0.09 via magneto-optics[J]. Physical Review B, 2012, 85(19): 195440.
\bibitem{Neupane M}Neupane M, Xu S Y, Sankar R, et al. Observation of a three-dimensional topological Dirac semimetal phase in high-mobility Cd 3 As 2[J]. Nature communications, 2014, 5: 3786.
\bibitem{Liu Z K}Liu Z K, Zhou B, Zhang Y, et al. Discovery of a three-dimensional topological Dirac semimetal, Na3Bi[J]. Science, 2014, 343(6173): 864-867.
\bibitem{Kashuba A B}Kashuba A B. Conductivity of defectless graphene[J]. Physical Review B, 2008, 78(8): 085415.
\bibitem{Wu C Hcurrent}Wu C H. Dynamical current-current correlation in the two-dimensional parabolic Dirac system[J]. arXiv preprint arXiv:1810.02413, 2018.
\bibitem{Moll P J W}Moll P J W, Potter A C, Nair N L, et al. Magnetic torque anomaly in the quantum limit of Weyl semimetals[J]. Nature communications, 2016, 7: 12492.
\bibitem{Zhou B}Zhou B, Lu H Z, Chu R L, et al. Finite size effects on helical edge states in a quantum spin-Hall system[J]. Physical review letters, 2008, 101(24): 246807.
\bibitem{WuElectronic transport}Wu C H. Electronic transport and the related anomalous effects in silicene-like hexagonal lattice[J]. arXiv preprint arXiv:1807.10898, 2018.

\bibitem{Wu C Hrkky}Wu C H. Two-dimensional parabolic Dirac system in the presence of non-magnetic and magnetic impurities[J]. arXiv preprint arXiv:1809.09289, 2018.
\bibitem{Hosseini M V}Hosseini M V, Askari M. Ruderman-Kittel-Kasuya-Yosida interaction in Weyl semimetals[J]. Physical Review B, 2015, 92(22): 224435.
\bibitem{Zare M}Zare M, Parhizgar F, Asgari R. Strongly anisotropic RKKY interaction in monolayer black phosphorus[J]. Journal of Magnetism and Magnetic Materials, 2018, 456: 307-315.
\bibitem{Chang H R}Chang H R, Zhou J, Wang S X, et al. RKKY interaction of magnetic impurities in Dirac and Weyl semimetals[J]. Physical Review B, 2015, 92(24): 241103.


\bibitem{Saito R}Saito R, Kamimura H. Orbital susceptibility of higher-stage graphite intercalation compounds[J]. Physical Review B, 1986, 33(10): 7218.
\bibitem{Ando T}Ando T. Exotic electronic and transport properties of graphene[J]. Physica E: Low-dimensional Systems and Nanostructures, 2007, 40(2): 213-227.
\bibitem{Scriba J}Scriba J, Wixforth A, Kotthaus J P, et al. The effect of Landau quantization on cyclotron resonance in a non-parabolic quantum wells[J]. Semiconductor Science and Technology, 1993, 8(1S): S133.
\bibitem{Shakouri K2}Shakouri K, Vasilopoulos P, Vargiamidis V, et al. Spin-and valley-dependent commensurability oscillations and electric-field-induced quantum Hall plateaux in periodically modulated silicene[J]. Applied Physics Letters, 2014, 104(21): 213109.
\bibitem{Sigg H}Sigg H, Perenboom J, Pfeffer P, et al. Non-parabolicity and anisotropy in the conduction band of GaAs[J]. Solid state communications, 1987, 61(11): 685-689.
\bibitem{Wu C Hinteger}Wu C H. Integer quantum Hall conductivity and longitudinal conductivity in silicene under the electric field and magnetic field[J]. arXiv preprint arXiv:1805.10656, 2018.
\bibitem{Tahir M}Tahir M, Schwingenschl{\"o}gl U. Valley polarized quantum Hall effect and topological insulator phase transitions in silicene[J]. Scientific reports, 2013, 3: 1075.
\bibitem{Tahir M2}Tahir M, Manchon A, Schwingenschl{\"o}gl U. Photoinduced quantum spin and valley Hall effects, and orbital magnetization in monolayer MoS 2[J]. Physical Review B, 2014, 90(12): 125438.
\bibitem{Vargiamidis V}Vargiamidis V, Vasilopoulos P, Hai G Q. Dc and ac transport in silicene[J]. Journal of Physics: Condensed Matter, 2014, 26(34): 345303.
\bibitem{Thakur A}Thakur A, Sadhukhan K, Agarwal A. Dynamic current-current susceptibility in three-dimensional Dirac and Weyl semimetals[J]. Physical Review B, 2018, 97(3): 035403.
\bibitem{Wu C HTight}Wu C H. Tight-binding model and ab initio calculation of silicene with strong spin-orbit coupling in low-energy limit[J]. arXiv preprint arXiv:1804.01695, 2018.
\bibitem{Xu N}Xu N, Chen Q, Tian H, et al. Diamagnetic response in zigzag hexagonal silicene rings[J]. Physics Letters A, 2016, 380(39): 3229-3232.
\bibitem{Koshino M2}Koshino M, Ando T. Diamagnetism in disordered graphene[J]. Physical Review B, 2007, 75(23): 235333.
\bibitem{Goswami P}Goswami P, Chakravarty S. Quantum criticality between topological and band insulators in 3+ 1 dimensions[J]. Physical review letters, 2011, 107(19): 196803.
\bibitem{Moors K}Moors K, Zyuzin A A, Zyuzin A Y, et al. Disorder-driven exceptional lines and Fermi ribbons in tilted nodal-line semimetals[J]. arXiv preprint arXiv:1810.03191, 2018.
\bibitem{Altshuler B L}Altshuler B L, Aronov A G, Lee P A. Interaction effects in disordered Fermi systems in two dimensions[J]. Physical Review Letters, 1980, 44(19): 1288.
\bibitem{Shiranzaei M}Shiranzaei M, Fransson J, Cheraghchi H, et al. Nonlinear spin susceptibility in topological insulators[J]. Physical Review B, 2018, 97(18): 180402.
\bibitem{Vollhardt D}Vollhardt D, W{\"o}lfle P. Diagrammatic, self-consistent treatment of the Anderson localization problem in d≤ 2 dimensions[J]. Physical Review B, 1980, 22(10): 4666.
\bibitem{Wu C HElectronic}Wu C H. Electronic transport and dynamical polarization in bilayer silicene-like system[J]. arXiv preprint arXiv:1809.05983, 2018.
\bibitem{Park S}Park S, Min H, Hwang E H, et al. Diluted magnetic Dirac-Weyl materials: Susceptibility and ferromagnetism in three-dimensional chiral gapless semimetals[J]. arXiv preprint arXiv:1804.10867, 2018.
\bibitem{Liu Y}Liu Y, Ruden P P. Temperature-dependent anisotropic charge-carrier mobility limited by ionized impurity scattering in thin-layer black phosphorus[J]. Physical Review B, 2017, 95(16): 165446.
\bibitem{Wu C HDynamical}Wu C H. Dynamical polarization, plasmon model, and the Friedel oscillation of the screened potential in doped Dirac and Weyl system[J]. arXiv preprint arXiv:1809.00169, 2018.
\bibitem{Wu C HRINP}Wu C H. Dynamical polarization and the optical response of silicene and related materials[J]. Results in Physics, 2018.

\bibitem{Wu C HInterband}Wu C H. Interband and intraband transition, dynamical polarization and screening of the monolayer and bilayer silicene in low-energy tight-binding model[J]. arXiv preprint arXiv:1805.07736, 2018.
\bibitem{Stano P}Stano P, Klinovaja J, Yacoby A, et al. Local spin susceptibilities of low-dimensional electron systems[J]. Physical Review B, 2013, 88(4): 045441.
\bibitem{Stauber T}Stauber T, Zimmermann R, Castella H. Electron-phonon interaction in quantum dots: A solvable model[J]. Physical Review B, 2000, 62(11): 7336.
\bibitem{Kumar V}Kumar V, Kumar U, Setlur G S. Quantum Rabi oscillations in graphene[J]. JOSA B, 2014, 31(3): 484-493.
\bibitem{Wu C HGeometrical}Wu C H. Geometrical structure and the electron transport properties of monolayer and bilayer silicene near the semimetal-insulator transition point in tight-binding model[J]. arXiv preprint arXiv:1805.00350, 2018.
\bibitem{Wu C HAnomalous}Wu C H. Anomalous Rabi oscillation and related dynamical polarizations under the off-resonance circularly polarized light[J]. arXiv preprint arXiv:1806.03592, 2018.
\bibitem{Sodemann I}Sodemann I, Fogler M M. Interaction corrections to the polarization function of graphene[J]. Physical Review B, 2012, 86(11): 115408.
\bibitem{Vafek O}Vafek O, Case M J. Renormalization group approach to two-dimensional Coulomb interacting Dirac fermions with random gauge potential[J]. Physical Review B, 2008, 77(3): 033410.



\bibitem{Wu C Htime}Wu C H. Time Evolution and Thermodynamics for the Nonequilibrium System in Phase-Space[J]. Canadian Journal of Physics, 2018 (ja).
\bibitem{Mishchenko E G}Mishchenko E G. Effect of electron-electron interactions on the conductivity of clean graphene[J]. Physical review letters, 2007, 98(21): 216801.



\end{thebibliography}
\end{document}